\documentclass[prl,twocolumn,showpacs,superscriptaddress]{revtex4-1} 
\usepackage{graphicx}
\usepackage{amsmath}
\usepackage{amssymb}
\usepackage{bm}
\usepackage{dsfont}
\usepackage{hyperref}
\usepackage{multirow}
\usepackage{appendix}
\usepackage{color}
\usepackage{psfrag}
\usepackage{epsfig}
\usepackage[tight]{subfigure}

\begin{document}

\title{Detection of Majorana Kramers pairs using a quantum point contact}
\author{Jian Li}
\affiliation{
Department of Physics,
Princeton University,
Princeton, NJ 08544, USA
            }

\author{Wei Pan}
\affiliation{
Sandia National Laboratories,
Albuquerque, NM 87185, USA
            }

\author{B. Andrei Bernevig}
\affiliation{
Department of Physics,
Princeton University,
Princeton, NJ 08544, USA
            }

\author{Roman M. Lutchyn}
\affiliation{
Station Q,
Microsoft Research,
Santa Barbara, California 93106-6105, USA
            }
\date{\today}
\begin{abstract}
  We propose a setup that integrates a quantum point contact (QPC) and a Josephson junction on a quantum spin Hall sample, experimentally realizable in InAs/GaSb quantum wells. The confinement due to both the QPC and the superconductor results in a Kramers pair of Majorana zero-energy bound states when the superconducting phases in the two arms differ by an odd multiple of $\pi$ across the Josephson junction. We investigate the detection of these Majorana pairs with the integrated QPC, and find a robust switching from normal to Andreev scattering across the edges due to the presence of Majorana Kramers pairs. This transport signature is expected to be exhibited in measurements of differential conductance and/or current cross-correlation at low bias.
\end{abstract}
\maketitle

In the pursuit of Majorana zero-energy modes~\cite{wilczek_majorana_2009, alicea_new_2012, beenakker_search_2013, stanescu_majorana_2013, elliott_colloquium:_2014, NayakReview2015} in topological superconductors several experimental and theoretical directions are being explored. Most noticeably, these include topological insulator/superconductor structures \cite{fu_superconducting_2008, fu_josephson_2009, nilsson_splitting_2008, knez_andreev_2012, yu_superconducting_2014}, semiconductor-superconductor heterostructures~\cite{Sau2010, lutchyn_majorana_2010, oreg_helical_2010, cook_majorana_2011, mourik_signatures_2012, Rokhinson2012, das_zero-bias_2012, Deng2012, Fink2012, Churchill2013, Deng_arxiv2014, Krogstrup2015}, and magnetically-ordered metallic systems coupled to an s-wave superconductor~\cite{choy_majorana_2011, martin_majorana_2012, nadj-perge_proposal_2013, klinovaja_topological_2013, braunecker_interplay_2013, vazifeh_self-organized_2013, pientka_topological_2013, li_majorana_2014, nadj-perge_observation_2014}. In all cases, proximity to superconductivity (SC) leads to the formation of the Majorana zero-energy bound states localized at the defects such as vortices and domain walls. Such defects have been predicted to obey non-Abelian braiding statistics~\cite{Moore1991, Nayak1996, BondersonNonAbelianStatistics}, and, as such, might be useful for topological quantum computation~\cite{kitaev_unpaired_2001, TQCreview, AliceaBraiding}. The detection of MBSs often involves measurements of transport signatures such as zero-bias anomalies \cite{law_majorana_2009, flensberg_tunneling_2010, fidkowski_universal_2012, mourik_signatures_2012, das_zero-bias_2012}, quantized conductance \cite{law_majorana_2009, akhmerov_quantized_2011, wimmer_quantum_2011}, or fractional Josephson effect \cite{kitaev_unpaired_2001, kwon_fractional_2004, fu_josephson_2009, lutchyn_majorana_2010, Cheng2015}.  Most of these signatures, however, can be obscured by real-world problems like disorder \cite{liu_zero-bias_2012, bagrets_class_2012, pikulin_zero-voltage_2012} and quasi-particle poisoning \cite{mannik_effect_2004, aumentado_nonequilibrium_2004}, making the detection of MBSs challenging.

\begin{figure}
  \centering
  \includegraphics[width=0.4\textwidth]{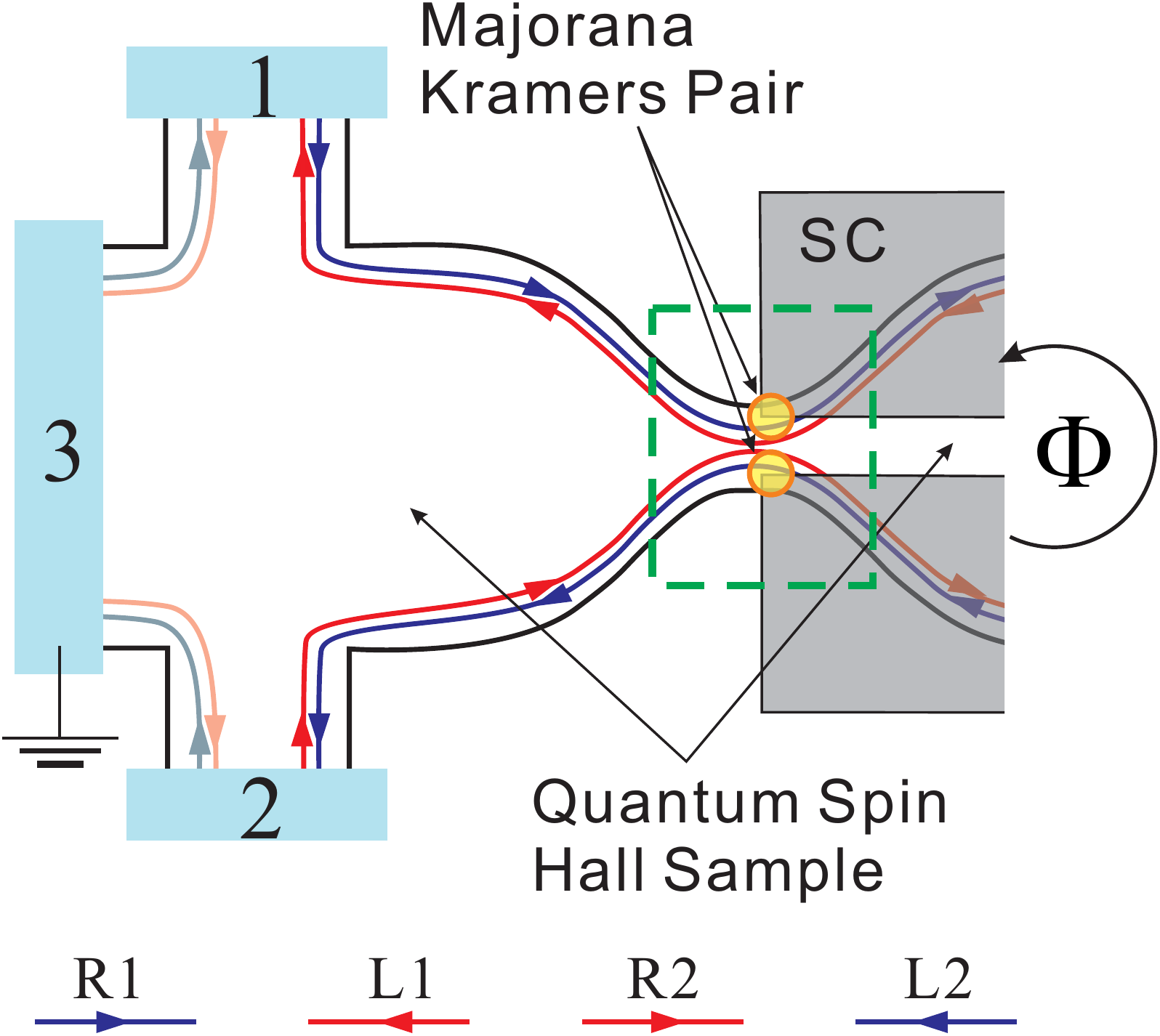}
  \caption{Our proposed setup: a quantum point contact build in a quantum spin Hall sample is half covered by a Josephson junction; three metallic contacts of a (half) Hall-bar configuration are away from the point contact, with contact 3 always grounded.\label{fig:setup}}
\end{figure}

In this paper, we propose a setup for preparing and observing MBSs in systems exhibiting quantum spin Hall (QSH) effect \cite{Kane:2005, bernevig06, konig07}, such as HgTe quantum wells~\cite{bernevig06, konig07, roth_nonlocal_2009}, and InAs/GaSb quantum wells \cite{liu_quantum_2008, knez_evidence_2011, knez_andreev_2012, du_observation_2013, yu_superconducting_2014}. The setup, illustrated in Fig.~\ref{fig:setup}, has a built-in quantum point contact (QPC) with a conventional superconductor (SC) junction covering a half of the constriction. The two parts of the SC are connected far away from the QPC region, and consequently the SC phase difference $\varphi$ across the SC junction can be tuned by a weak perpendicular magnetic field.
A pair of zero-energy MBSs is localized at the junction when $\varphi$ is an odd multiple of $\pi$ \cite{fu_josephson_2009}. At these points, time-reversal symmetry (TRS) applies approximately as long as the magnetic field that creates the phase difference for the SC arms is weak enough, and the MBSs form a Kramers pair. Using the scattering approach for noninteracting electrons \cite{buttiker_four-terminal_1986, Buttiker1992}, we analyze the transport signatures related to these MBSs and propose several ways to measure them using the QPC. The advantage of this proposal is that several key steps such as the experimental realization of a QPC, as well as proximity induced pairing in the edge states, in an InAs/GaSb QSH sample have been already performed \cite{shi_giant_2014}.

We start by analytically finding the bound-state solutions at the interface between the quantum point contact and the SC (see Fig.~\ref{fig:setup} highlighted region). In the interface region, the effective Hamiltonian is given by
\begin{align}\label{eq:ham}
  \hspace{-3mm}
  H = \int dx &\sum\limits_{n=1,2} \Bigl[\psi_{Rn}^\dagger(\hbar v_F\hat{k} - \mu)\psi_{Rn} + \psi_{Ln}^\dagger(-\hbar v_F\hat{k} - \mu)\psi_{Ln}\Bigr] \nonumber\\
  &+\sum\limits_{n=1,2}\Delta_n(x) e^{i\varphi_n(x)} \psi_{Rn}^\dagger\psi_{Ln}^\dagger + h.c. \nonumber\\
  &+m(x)(\psi_{R1}^\dagger\psi_{L2}+\psi_{L1}^\dagger\psi_{R2})+h.c. \nonumber\\
  &+f(x)(\psi_{R1}^\dagger\psi_{R2}-\psi_{L1}^\dagger\psi_{L2})+h.c.,
\end{align}
where $\psi_{R/Ln} \equiv \psi_{R/Ln}(x)$ is the fermionic field (annihilation) operator near the Fermi energy for the right/left moving edge states along the upper ($n=1$) or lower ($n=2$) edge, $v_F$ is the Fermi velocity, $\hat{k} \equiv -i\partial_x$, and all $\Delta_n(x)$, $m(x)$ and $f(x)$ are real without loss of generality. Physically, $m$ and $f$ represent the hybridization gaps
possible when the two edges of the QSH sample approach near the QPC; $\Delta$ is the induced SC gap with phases $\varphi_1$ and $\varphi_2$ in different arms (see Fig.~\ref{fig:setup}). This Hamiltonian transforms under time-reversal and particle-hole symmetries (PHS) as follows:
\begin{align}
  &TH(\varphi_1,\varphi_2)T^{-1} = H(-\varphi_1,-\varphi_2), \label{eq:Tham}\\
  &PH(\varphi_1,\varphi_2)P^{-1} = -H(\varphi_1,\varphi_2). \label{eq:Pham}
\end{align}
Here, by definition, $T\psi_{Rn}T^{-1} = s_n\psi_{Ln}$, $T\psi_{Ln}T^{-1} = -s_n \psi_{Rn}$ with $s_{1}$=$1$ and $s_{2}$=$-1$, $P\psi_{R/Ln}P^{-1} = \psi_{R/Ln}^\dagger$, $TaT^{-1} = a^*$ and $PaP^{-1} = a^*$ if $a$ is a number. For simplicity, we further assume $\Delta_n(x) = s_n\theta(x)\Delta$, $m(x) = \theta(-x)m$, $f(x) = \theta(-x)f$ with $\theta(x)$ the Heaviside step function and $\Delta,m,f>0$, as well as
\begin{align}
  \varphi_1(x) = (1 - \epsilon_0 x)\varphi,\quad \varphi_2(x) = (\epsilon_0 x)\varphi \quad (\epsilon_0\ge 0).
\end{align}
Physically, $\epsilon_0$ is roughly proportional to the inverse of the circumference of the SC that encloses the magnetic flux. We assume $\epsilon_0$ to be sufficiently small such that both $\varphi_1$ and $\varphi_2$ are slowly varying at the length scale of the SC coherence length $\xi = \hbar v_F/\Delta$.
If $\epsilon_0 \neq 0$, there is a finite TRS-breaking splitting energy between two MBSs $\delta E = \hbar v_F\epsilon_0\varphi/2$ even if $\varphi$ is an integer multiple of $\pi$ [cf. Eq.~\eqref{eq:Tham}].

\begin{figure}
  \centering
  \includegraphics[width=0.45\textwidth]{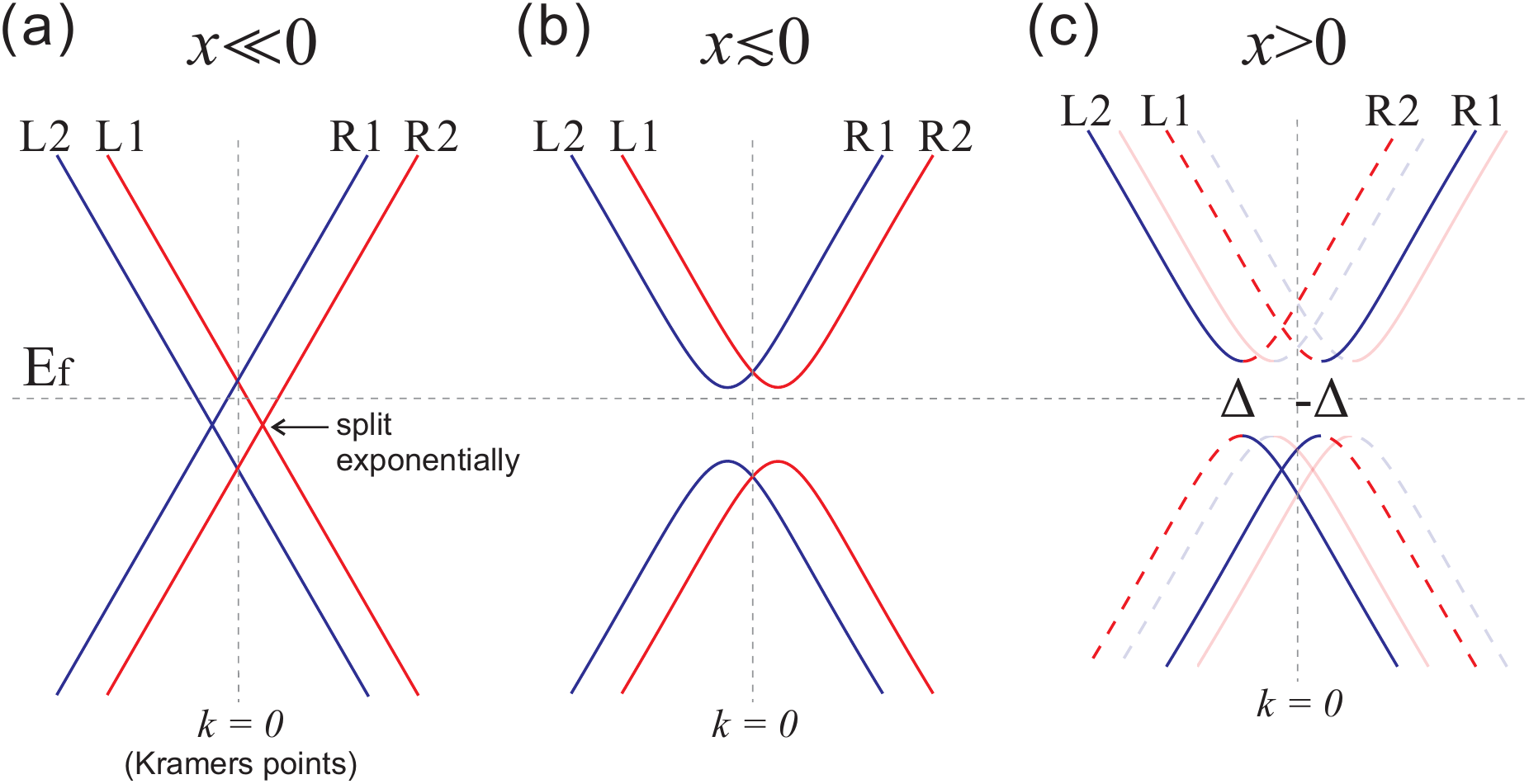}
  \caption{Schematic explanation of the origin of the Majorana Kramers pairs.\label{fig:bands}}
\end{figure}

The existence of a Kramers pair of MBSs in the limit of $\epsilon_0=0$ is based on a simple physical argument. In Fig.~\ref{fig:bands}a we present the system's every dispersion in the normal region far from the QPC ($x\ll 0$), featured by the two gapless one-dimensional Dirac dispersions of the lower and upper QSH edges. As the edges get closed together near the QPC ($x\lesssim 0$), they hybridize and open a trivial gap at non-Kramers points (Fig.~\ref{fig:bands}b). On the SC side ($x>0$), a SC gap (larger than the hybridization gap, which is always the case when $x\gg 0$) is present at the Fermi level (which can be tuned by a gate) to create another insulator. This is a gapped superconductor, but, it is a topological one (Fig.~\ref{fig:bands}c). Notice that $R_1$, $R_2$ are right movers giving two $k>0$ Fermi points on the upper and lower edges, as such, the superconducting gaps differ by a sign on the two $k>0$ Fermi points when there is a $\pi$ phase shift in the two arms of the superconductor. Thus, at $x>0$ we have a time-reversal topological superconductor. A Kramers pair of MBS appears when the topological superconductor is put next to a trivial insulator $x\lesssim 0$. These qualitative arguments agree well with the analytic solution discussed below.

\begin{figure}
  \centering
  \includegraphics[width=0.4\textwidth]{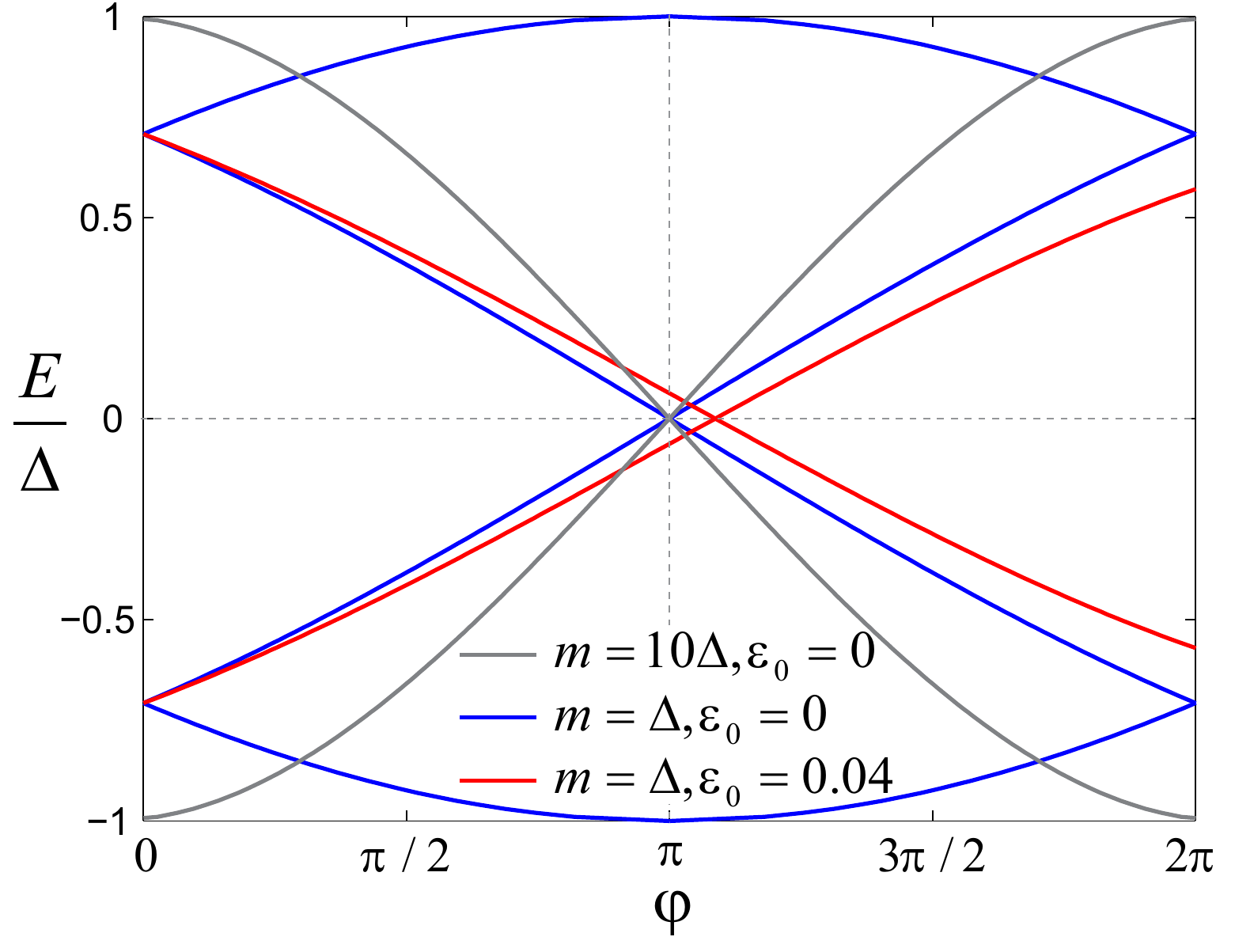}
  \caption{Spectra of the subgap states obtained numerically from Eq.~\eqref{eq:sp0}. Cases are compared between strong (grey line) and mild (blue line) point contact constriction, as well as between constant (blue line) and spatially varying (red line) pairing phases.
  \label{fig:spectrum}}
\end{figure}

In the subgap regime, namely $|E|<\min(m-|\mu|,\Delta-|\delta E|)$, the solutions of bound states can be obtained by matching eigenstate wavefunctions at the interface (see Supplemental Material Sec.~I.C). The eigenvalues  are determined by the following equation
\begin{align}\label{eq:sp}
  [e^{i(\alpha-\beta-2\gamma_-)}-e^{i\varphi}][e^{i(\beta-\alpha-2\gamma_+)}-e^{i\varphi}]=0,
\end{align}
where $\cos\alpha = (\mu+E)/m$, $\cos\beta = (\mu-E)/m$, $\cos\gamma_\pm = (\delta E \pm E)/\Delta$ with $\alpha, \beta, \gamma_\pm\in(0,\pi)$. Eq.~\eqref{eq:sp} is particle-hole symmetric, as changing the sign of $E$ exchanges the two terms on its left-hand side. In the limit of pinched-off QPC ($m\rightarrow\infty$), we obtain $E = \pm (\Delta \cos\varphi/2 \pm \delta E)$ \cite{kwon_fractional_2004, fu_josephson_2009}. The full subgap spectra in generic cases can be solved numerically, as exemplified in Fig.~\ref{fig:spectrum}.

Zero-energy solutions should satisfy $\alpha = \beta$ and $\gamma_+=\gamma_-=\gamma = \arccos(\delta E/\Delta) \in (0,\pi)$, hence Eq.~\eqref{eq:sp} is reduced to
\begin{align}\label{eq:sp0}
  e^{i(\varphi+2\gamma)} = 1.
\end{align}
When $\epsilon_0=0$, the zero energy solutions, occurring iff $\varphi = (2l+1)\pi$ with $l$ being an integer, define two Majorana operators
\begin{align}
  &\chi_1 = \int dx \bigl[(\tilde{\psi}_{R1}^{(-)} + i\tilde{\psi}_{L1}^{(+)} + i\tilde{\psi}_{R2}^{(+)} + \tilde{\psi}_{L2}^{(-)}) + h.c.\bigr], \\
  &\chi_2 = \int dx \bigl[(i\tilde{\psi}_{R1}^{(+)} + \tilde{\psi}_{L1}^{(-)} + \tilde{\psi}_{R2}^{(-)} + i\tilde{\psi}_{L2}^{(+)}) + h.c.\bigr],
\end{align}
where $\tilde{\psi}_{R/Ln}^{(\pm)} = [\theta(-x)e^{(m\sin\alpha) x}(\cos fx\pm\sin fx) + \theta(x)e^{-\Delta x}e^{s_{R/L}i\mu x}]\psi_{R/Ln}$ with $s_{R/L}=\mp 1$ (here we set $\hbar v_F=1$). As a result of TRS in this idealized case, $\chi_1$ and $\chi_2$ form a Kramers pair: $T\chi_1 T^{-1} = \chi_2$ and $T\chi_2 T^{-1} = -\chi_1$ \footnote{In fact, the strict presence of TRS will imply the existence of another pair of MBSs at the opposite side of the Josephson junction. They are of no concern to us in terms of the local measurements that we propose.}. When $\epsilon_0$ is finite but small (i.e. $|\delta E|/\Delta\ll 1$), Eq.~\eqref{eq:sp0} yields $\varphi = (2l+1)\pi/(1-\epsilon_0\xi)$. The zero-energy solutions shift to $\varphi$'s different from $(2l+1)\pi$, owing to the finite penetration of the bound states into the gapped SC, and are no longer exact Kramers partners (see Supplemental Material Sec.~I.D). We will neglect this subtlety in the following by assuming $\epsilon_0$ to be sufficiently small.


 Having established the presence of the MBSs at the QPC-SC interface, we now discuss their detection. To this end, the incorporation of the QPC in our proposed setup was particularly useful as the QSH edge states reflected at the QPC naturally become probes of the MBSs. In this case, the low-energy ($E<\Delta$) effective Hamiltonian in the presence of a Majorana Kramers pair, constrained by TRS, has a generic form
\begin{align}\label{eq:ham2}
  &H_{\pi} =-i\int_{-\infty}^{+\infty} d\tilde{x} \;\Bigl\{ \psi_{+}^\dagger\partial_{\tilde{x}}\psi_{+} - \psi_{-}^\dagger\partial_{\tilde{x}}\psi_{-}
  + \delta(\tilde{x})\bigl[ \nonumber\\
  &\chi_1(t_+\psi_{+}+t_-\psi_{-}+h.c.) + \chi_2(t_-^*\psi_{+}-t_+^*\psi_{-}+h.c.)\bigr] \Bigr\},
\end{align}
where $\tilde{x}$ stands for the unfolded coordinate such that $\psi_{+}(\tilde{x}) \equiv \theta(-\tilde{x})\psi_{R1}(\tilde{x})+\theta(\tilde{x})\psi_{L2}(-\tilde{x})$ and $\psi_{-}(\tilde{x}) \equiv \theta(-\tilde{x})\psi_{R2}(\tilde{x})+\theta(\tilde{x})\psi_{L1}(-\tilde{x})$; $t_{\pm}$ stands for the coupling between the MBSs and $\psi_{\pm}$. $H_\pi$ is manifestly time-reversal symmetric. The scattering matrix that relates the current amplitudes of the outgoing (electron and hole) components of $\psi_{\pm}$ to those of the incoming components can be obtained by using the formula \cite{nilsson_splitting_2008}
\begin{align}
  &S_{\pi}(E) = \mathds{1} -iW^\dag(E+\frac{i}{2}WW^\dag)^{-1}W, \label{eq:Sformula}\\
  &W = -i
  \begin{pmatrix}
    t_+ & t_- & t_+^* & t_-^* \\
    t_-^* & -t_+^* & t_- & -t_+ \\
  \end{pmatrix},
\end{align}
where $W$ is the coupling matrix between the scattering modes and the Majorana pair. This yields
\begin{align}\label{eq:Spi}
  S_{\pi}(E) = \frac{1}{iE-\Gamma}
  \begin{pmatrix}
    0 & iE & -A & C \\
    iE & 0 & C^* & -A \\
    A & C^* & 0 & iE \\
    C & A & iE & 0 \\
  \end{pmatrix},
\end{align}
where $\Gamma = |t_+|^2+|t_-|^2$, $A = t_+t_--(t_+t_-)^*$, $C = t_+^2+(t_-^2)^*$. We have chosen the outgoing basis in Eq.~\eqref{eq:Spi} so that both the incoming and the outgoing bases are ordered as (1e, 2e, 1h, 2h), where 1(2) stands for the upper(lower) arm and e(h) stands for the electron(hole) component of the original edge channels (cf. Fig.~\ref{fig:setup}). We immediately see that at $E=0$, all normal scattering processes, corresponding to the diagonal blocks in Eq.~\eqref{eq:Spi}, vanish; only local ($A$) and crossed ($C$) Andreev scatterings remain. This scenario represents a fixed point for the scattering corresponding to the presence of a Majorana Kramers pair.

Away from this fixed point, the consequence of lifted degeneracy at zero-energy can be investigated perturbatively by including an additional term $H_M(\varphi) = iE_{\varphi}\chi_1\chi_2$ ($E_\pi=0$) into Hamiltonian \eqref{eq:ham2}. In terms of the scattering matrix, this amounts to replacing $E$ on the right-hand side of Eq.~\eqref{eq:Sformula} by $E+E_{\varphi}\sigma_y$ with $\sigma_y$ the Pauli matrix. To the lowest order in $E_\varphi$, the correction to the scattering matrix in Eq.~\eqref{eq:Spi} is given by
\begin{align}\label{eq:Scorr}
  \delta S(E) = \frac{E_{\varphi}}{(iE-\Gamma)^2}
  \begin{pmatrix}
    -C & -A & -\Gamma & 0 \\
    A & C^* & 0 & \Gamma \\
    -\Gamma & 0 & -C^* & A \\
    0 & \Gamma & -A & C \\
  \end{pmatrix},
\end{align}
which suggests a suppression of crossed Andreev reflection and simultaneously an enhancement of normal scattering processes when $E_{\varphi}$ becomes large compared to $\max(|E|,\Gamma)$. Indeed this situation corresponds to another (trivial) fixed point as shown next through symmetry analysis.

In order to understand the scattering of helical edge states at the QPC-SC interface in a more general setting, we go back to the original Hamiltonian \eqref{eq:ham} and define the scattering matrix generically as \cite{fisher_relation_1981}
\begin{align}
  S_{n'\nu',n\nu}(E) = i\hbar v_F G_{n'\nu',n\nu}^R(x'_0,x_0;E),
\end{align}
where $n,n'=1,2$ stand for the lateral edges, $\nu,\nu'=e,h$ stand for electron or hole channels, and the retarded Green's functions $G_{n'\nu',n\nu}^R(x'_0,x_0;E) = \frac{1}{i\hbar}\int dt\, e^{iEt/\hbar} \theta(t) \bigl\langle\{ \psi_{Ln'\nu'}(x'_0,t), \psi_{Rn\nu}^{\dagger}(x_0,0) \} \bigr\rangle$ with $\psi_{e} = \psi$ and $\psi_{h} = \psi^\dagger$ in terms of the original field operators in Hamiltonian \eqref{eq:ham}. Both $x'_0$ and $x_0$ are chosen far away from the QPC so that the scattering channels are well-defined; the explicit choice of $x'_0$ and $x_0$ is otherwise not important.

The scattering matrices are constrained by PHS and TRS, respectively, as (see Supplemental Material Sec.~II.A):
\begin{align}
  &S_{n'\nu',n\nu}(E,\varphi_{1,2}) = S_{n'\bar{\nu'},n\bar{\nu}}(-E,\varphi_{1,2})^*, \label{eq:SbyPHS}\\
  &S_{n'\nu',n\nu}(E,\varphi_{1,2}) =  -s_ns_{n'} S_{n\nu,n'\nu'}(E,-\varphi_{1,2}), \label{eq:SbyTRS}
\end{align}
where $\bar{e}=h$, $\bar{h}=e$, and $s_n$ is defined below Eq.~\eqref{eq:Pham}. Eqs.~\eqref{eq:SbyPHS} and \eqref{eq:SbyTRS} together imply that, at $E\simeq 0$ and $\varphi\simeq l\pi$, $S$ takes the form
\begin{align}\label{eq:S0}
  S_{0,\pi}(E\simeq 0) =
  \begin{pmatrix}
    0 & b & -ia_1 & c \\
    b & 0 & c^* & -ia_2 \\
    ia_1 & c^* & 0 & b^* \\
    c & ia_2 & b^* & 0
  \end{pmatrix},
\end{align}
in the same basis (1e, 2e, 1h, 2h) as in Eq.~\eqref{eq:Spi}. Here, $a_1$ and $a_2$, both real, stand for local Andreev reflections involving either edge 1 or 2; $b$ and $c$ stand for normal back-scattering and crossed Andreev reflection, respectively, from one edge to the other. Note that normal back-scattering within one edge is forbidden by TRS (diagonal terms vanish). By further using the unitarity condition, $S$ is limited down to two possibilities (see Supplemental Material Sec.~II.A):
\begin{align}
  &\varphi\simeq 2l\pi\,: &c = 0, b \ne 0, a_1=-a_2\,; \label{eq:phieven} \\
  &\varphi\simeq (2l+1)\pi\,: &b = 0, c \ne 0, a_1=a_2\,. \label{eq:phiodd}
\end{align}
The latter case reproduces Eq.~\eqref{eq:Spi} by identifying $a_1=a_2=iA/\Gamma$ and $c=-C/\Gamma$. Physically, the above equations imply that the zero-energy scattering processes at the QPC-SC interface will be entirely Andreev reflections, local and crossed, in the presence of MBSs at $\varphi\simeq (2l+1)\pi$; the crossed Andreev reflection probability will be gradually suppressed to zero when $\varphi$ is tuned towards $2l\pi$, meanwhile normal scattering between edges will become finite. This is again consistent with our previous result Eq.~\eqref{eq:Scorr} based on perturbation to the effective Hamiltonian \eqref{eq:ham2}.

\begin{figure}
  \centering
  \includegraphics[width=0.46\textwidth]{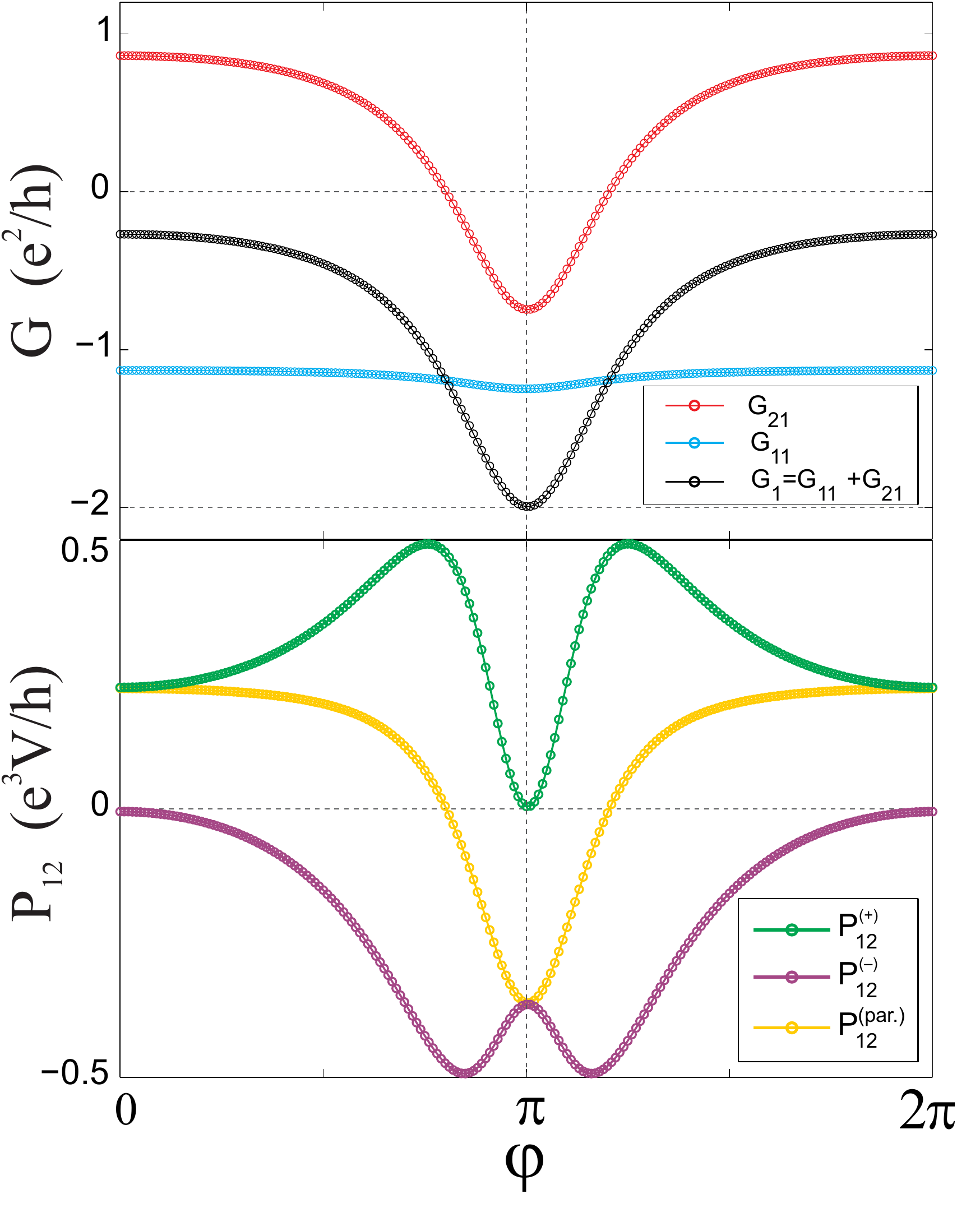}
  \caption{Simulation results of two types of measurements on our proposed setup: differential conductances (upper panel) and zero-frequency current cross-correlators (lower panel), obtained numerically by using the Bogoliubov-de Gennes form of the Bernevig-Hughes-Zhang Hamiltonian \cite{bernevig06}. These results verify and complement the features analyzed by using the effective edge theory (see main text).\label{fig:sim}}
\end{figure}

The switching between normal backscattering and crossed Andreev reflection will lead to significant effects in transport measurements. At zero temperature, the zero-bias differential conductances, defined by $G_{mn} = dI_m/dV_n$ will exhibit this switching behavior. In particular, we find
\begin{align}
  &G_{21} = (|b|^2-|c|^2)e^2/h, \label{eq:G21}\\
  &G_{1} = G_{11}+G_{21} = -2(|a_1|^2+|c|^2)e^2/h, \label{eq:G1}
\end{align}
where $a_{1}$, $b$ and $c$ are subject to the scattering matrix normalization condition. Both $G_{21}$ and $G_{1}$ can be measured by biasing only contact 1 and grounding contacts 2 and 3. \footnote{To be more precise, for the setup we are considering, Eq.~\eqref{eq:G1} must also include contact 3, such that $G_{1} = G_{11}+G_{21}+G_{31}$. However, because the current flow from 1 to 3 is assumed to be always trivial: $G_{31} = e^2/h$ regardless of $\varphi$, we have effectively defined $G_{11}$ as $G_{11}+G_{31}$.} The above expressions are valid even in the absence of TRS, by interpreting $ia_1$ as $S_{1h,1e}$, $b$ as $S_{2e,1e}$, and $c$ as $S_{2h,1e}$ in general. When $\varphi$ is tuned from $2l\pi$ to $(2l+1)\pi$, the sign of $G_{21}$ flips, whereas the magnitude of $G_1$ reaches a quantized peak of magnitude $2e^2/h$ (see Fig.~\ref{fig:sim} upper panel).

Another quantity of interest is the current-current correlation function. Here, we focus on the zero-frequency current cross-correlation function between contacts 1 and 2 (while contact 3 is always grounded), defined by $P_{12} = \int_{-\infty}^{\infty}dt\, \frac{1}{2} \langle\{ \delta\hat{I}_1(t),\delta\hat{I}_2(0)\}\rangle$ with current fluctuation $\delta\hat{I}_{n = 1,2}(t) = \hat{I}_n(t) - \langle \hat{I}_{n} \rangle$. Within the scattering approach \cite{Buttiker1992, anantram_current_1996, Blanter2000, li_scattering_2012, edge_z_2_2013}, the scattering matrix in Eq.~\eqref{eq:S0} implies (see Supplemental Material Sec.~II.B):
\begin{align}
  P_{12}^{(+)} = 2\frac{e^3V}{h} a_{1}^2|b|^2,\quad
  P_{12}^{(-)} = -2\frac{e^3V}{h} a_{1}^2|c|^2,
\end{align}
where $P_{12}^{(+/-)}$ is the current cross-correlator measured with contacts 1 and 2 biased equally/oppositely. Although these formulas are valid only for low bias voltage at $\varphi\simeq l\pi$, they enable a straightforward examinations of the suppression of  normal backscattering ($P_{12}^{(+)}$) or crossed Andreev reflection ($P_{12}^{(-)}$). More generally, we find that the following relation holds for all $\varphi$ (see Supplemental Material Sec.~II.B and Fig.~\ref{fig:sim} lower panel):
\begin{align}\label{eq:Peq}
  P_{12}^{(+)}+P_{12}^{(-)}-P_{12}^{\text{(par.)}} = 0,
\end{align}
where $P_{12}^{\text{(par.)}} = -({e^3V}/{h})(G_{11}G_{21}+G_{12}G_{22})$ is conventionally called the partition noise \cite{Buttiker1992}. The partition noise is composed of multiplications of $G$'s, and hence exhibits the same flip of sign between $\varphi\simeq 2l\pi$ and $\varphi\simeq (2l+1)\pi$ (see Fig.~\ref{fig:sim} lower panel). The measurement of current cross-correlation function can, therefore, provide an additional check of the predicted switching between normal backscattering and crossed Andreev reflection.


 So far our analysis relied on one-dimensional effective edge theory and symmetry constraints. We further corroborate our conclusions by performing numerical simulations with the microscopic Bernevig-Hughes-Zhang Hamiltonian~\cite{bernevig06} in two dimensions (see Supplemental Material Sec.~II.C). These simulations include explicitly the metallic contacts in our proposed setup shown in Fig.~\ref{fig:setup}, and allow one to add various perturbations. Our main results, shown in Fig.~\ref{fig:sim}, agree very well with the predictions of Eqs.~(\ref{eq:phieven}-\ref{eq:Peq}). In addition, we verify that the switching from normal to Andreev scattering processes between two contacts (1 and 2) is a robust signature of the presence of the MBSs, which: i) remains valid when a finite $\epsilon_0$ is taken into account; ii) persists even when the QSH sample region covered by the SC is slightly doped; iii) vanishes if the region covered by the SC does not support any helical edge states (i.e. topologically trivial).

Finally, we have estimated the magnetic field and the temperature required to access the predicted transport signatures. From Ref.~\onlinecite{pribiag_edge-mode_2015}, the width of edge modes in InAs/GaSb quantum wells is estimated to be about 250 nm, therefore the device width in our proposal has to be larger than 500 nm. Combined with a device length about 1 $\mu$m, a magnetic field of 2 mT is needed to generate a half of the magnetic flux quantum. From Ref. \onlinecite{shi_giant_2014}, the energy gap of the proximity-effect induced superconductivity in InAs/GaSb quantum wells was observed to be around a few Kelvins, which is much larger than the base temperature (about 10 mK) that can be readily reached in a conventional dilution refrigerator. This suggests that the current low temperature technique is adequate for the proposed measurements of Majorana zero modes.


This work was supported by ONR-N00014-14-1-0330. RL acknowledges the hospitality of the Aspen Center for Physics supported by NSF grant No. PHY-1066293, where part of this work was done. BAB acknowledges support from NSF CAREER DMR-0952428, ONR-N00014-11-1-0635, MURI-130-6082, DARPA under SPAWAR Grant No. N66001-11-1-4110, the Packard Foundation, and the Keck grant. RL acknowledges the hospitality of the Aspen Center for Physics supported by NSF grant No. PHY–1066293, where part of this work was done. JL acknowledges support from Swiss National Science Foundation, ONR-N00014-11-1-0635, and MURI-130-6082. The work at Sandia was supported by was supported by the Department of Energy, Office of Basic Energy Sciences, Division of Materials Sciences and Engineering.

\bibliography{qpc_refs}

\clearpage
\newpage

\setcounter{equation}{0}
\renewcommand{\theequation}{S\arabic{equation}}
\setcounter{figure}{0}
\renewcommand{\thefigure}{S\arabic{figure}}

\appendix
\begin{widetext}
\begin{center}
\large
\textbf{
Supplemental material of ``Detection of Majorana Kramers pairs using a quantum point contact''}
\normalsize
\end{center}


\section{Bound state solutions}\label{sec:sol}

\subsection{Bogoliubov-de Gennes Hamiltonian}
To obtain the bound state solutions we write the interface Hamiltonian in the Nambu basis
\begin{align}
  &\mathcal{H}_{x<0} =
  \Psi^\dagger
  \begin{pmatrix}
    h_N(\hat{k})-\mu & 0 \\
    0 & -h_N(\hat{k})+\mu \\
  \end{pmatrix}\Psi, \label{eq:haml0}\\
  &\mathcal{H}_{x>0} =
  \tilde{\Psi}^\dagger
  \begin{pmatrix}
    h_{S1}(\hat{k}) & 0 & 0 & 0\\
    0 & h_{S2}(\hat{k}) & 0 & 0\\
    0 & 0 & h_{S1}(-\hat{k}) & 0 \\
    0 & 0 & 0 & h_{S2}(-\hat{k}) \\
  \end{pmatrix}\tilde{\Psi}, \label{eq:hamr0}
\end{align}
where $\hat{k}$ is the momentum (operator) parallel to the pinched edges,
\begin{align}
  &h_N(\hat{k}) =
  \begin{pmatrix}
    \hbar v_F\hat{k}\sigma_z & m\sigma_x+f\sigma_z \\
    m\sigma_x+f\sigma_z & \hbar v_F\hat{k}\sigma_z \\
  \end{pmatrix}, \\
  &\Psi = (\psi_{R1},\psi_{L1},\psi_{R2},\psi_{L2},\psi_{L1}^\dagger,-\psi_{R1}^\dagger,-\psi_{L2}^\dagger,\psi_{R2}^\dagger)^T, \label{eq:basis0}\\
  &h_{Sn}(\hat{k}) =
  \begin{pmatrix}
    \hbar v_F\hat{k}-\mu & \Delta_n \\
    \Delta_n^* & \mu-\hbar v_F\hat{k} \\
  \end{pmatrix}\quad (n=1,2),\\
  &\tilde{\Psi} = (\psi_{R1},\psi_{L1}^\dagger,\psi_{R2},-\psi_{L2}^\dagger,\psi_{L1},-\psi_{R1}^\dagger,\psi_{L2},\psi_{R2}^\dagger)^T.
\end{align}
$h_N$ is the effective Hamiltonian for two paris of hybridized QSH edge states with $m$ and $f$ the two possible gaps; $h_{Sn}$ is the Bogoliubov-de Gennes (BdG) Hamiltonian for a single proximity-induced s-wave pairing channel at the $n$-th edge; $\Psi$ and $\tilde{\Psi}$ are two different bases that are convenient for the normal side ($x<0$) and the superconducting side ($x>0$), respectively. We neglect all other pairing channels and assume $|\Delta_n|$ to be a constant independent of $\varphi$.
For simplicity we set $\hbar=v_F=1$, $\Delta_{1} = \Delta e^{i(\varphi-2\delta E x)}$, $\Delta_{2} = \Delta e^{2i\delta E x}$, and $\Delta,m,f>0$, which is consistent with the main text. Here, $\delta E \simeq  \hbar v_F\varphi/2L$ with $L$ the circumference of SC enclosing the magnetic flux.

\subsection{Symmetries}
In the single-particle basis defined according to $\Psi$ in Eq.~\eqref{eq:basis0}, the time-reversal operator reads
\begin{align}
  \mathcal{T} =
  \begin{pmatrix}
    i\sigma_y & & & \\
    & -i\sigma_y & & \\
    & & i\sigma_y & \\
    & & & -i\sigma_y \\
  \end{pmatrix}\mathcal{K},\quad
  \mathcal{T}^2=-1,
\end{align}
and the particle-hole conjugation operator reads
\begin{align}
  \mathcal{P} =
  \begin{pmatrix}
    & & -i\sigma_y & \\
    & & & i\sigma_y \\
    i\sigma_y & & & \\
    & -i\sigma_y & & \\
  \end{pmatrix}\mathcal{K},\quad
  \mathcal{P}^2=1.
\end{align}
Their actions on the single-particle Bloch Hamiltonians, corresponding to Eqs.~\eqref{eq:haml0} and \eqref{eq:hamr0}, lead to the following relations [cf. Eqs.~(2) and(3) in the main text; note that $\varphi_1=\varphi-2\delta E x$ and $\varphi_2=2\delta E x$]
\begin{align}
  &\mathcal{T}H(k,\varphi,\delta E)\mathcal{T}^{-1} =
  H(-k,-\varphi,-\delta E), \\
  &\mathcal{P}H(k,\varphi,\delta E)\mathcal{P}^{-1} =
  -H(-k,\varphi,\delta E).
\end{align}
The behavior of $\delta E$ under the symmetries stems from the fact that by definition $\delta E x$ acts as a phase.

\subsection{Subgap spectrum}
The Hamiltonian at the QPC on the non-superconducting side has a bulk gap of size $m$ determined by the backscattering, whereas at the SC side it has a pairing gap $\Delta$. We focus on energies below both gaps, namely
\begin{align}
  |E|<\min(m-\mu,\Delta-|\delta E|), \quad (0\le\mu<m).
\end{align}
The general solutions for a fixed $E$ in this regime are given as follows.

For $h_{Ne}(k) = h_N(k)-\mu$ (taking $k=-i\partial/\partial_x$, and $x$ negative but close to 0):
\begin{align}
  &\psi^{(Ne)}_{\pm}(x<0) =
  \begin{pmatrix}
    e^{-i\alpha/2} \\ \pm e^{i\alpha/2} \\
    \pm e^{-i\alpha/2} \\ e^{i\alpha/2} \\
  \end{pmatrix} e^{(m\sin\alpha \mp i f)x}, \\
  &\cos\alpha \equiv (\mu+E)/m, \quad \alpha\in(0,\pi).
\end{align}
For $h_{Nh}(k) = -h_N(k)+\mu$:
\begin{align}
  &\psi^{(Nh)}_{\pm}(x<0) =
  \begin{pmatrix}
    e^{-i\beta/2} \\ \pm e^{i\beta/2} \\
    \pm e^{-i\beta/2} \\ e^{i\beta/2} \\
  \end{pmatrix} e^{(m\sin\beta \mp i f)x}, \\
  &\cos\beta \equiv (\mu-E)/m, \quad \beta\in(0,\pi).
\end{align}
For $h_{S1}(\pm k)$ and $h_{S2}(\pm k)$, respectively:
\begin{align}
  &\psi^{(S1)}_{\pm}(x>0) =
  \begin{pmatrix}
    e^{i(\pm\mu-\delta E)x}e^{i(\varphi+\gamma_\pm)/2} \\ \pm e^{i(\pm\mu+\delta E)x}e^{-i(\varphi+\gamma_\pm)/2}
  \end{pmatrix} e^{-(\Delta\sin\gamma_\pm)x}\,, \label{eq:S1sol}\\
  &\psi^{(S2)}_{\pm}(x>0) =
  \begin{pmatrix}
    e^{i(\pm\mu+\delta E)x}e^{- i\gamma_\mp/2} \\ \mp e^{i(\pm\mu-\delta E)x}e^{i\gamma_\mp/2}
  \end{pmatrix} e^{-(\Delta\sin\gamma_\mp)x}, \\
  &\cos\gamma_\pm \equiv (\delta E \pm E)/\Delta, \quad \gamma_\pm\in(0,\pi).
\end{align}
These are solutions for the uncoupled blocks of the original Hamiltonian, in the bases defined by $\Psi$ and $\tilde{\Psi}$ for the normal and the superconducting parts, respectively, before the boundary condition is imposed.

From now on, we recover the eight-component form of the above solutions (by filling the irrelevant components with zeros) in the basis defined by $\Psi$ [see Eq.~\eqref{eq:basis0}]. One may verify the following symmetry relations (note that both $\mathcal{T}$ or $\mathcal{P}$ reverse the subscript sign):
\begin{align}
  & \mathcal{T}\psi^{(Ne)}_{+}(E) = \psi^{(Ne)}_{-}(E),\quad \mathcal{T}\psi^{(Nh)}_{+}(E) = \psi^{(Nh)}_{-}(E); \\
  &\mathcal{P}\psi^{(Ne)}_{+}(E) = \psi^{(Nh)}_{-}(-E),\quad \mathcal{P}\psi^{(Ne)}_{-}(E) = -\psi^{(Nh)}_{+}(-E); \\
  & \mathcal{T}\psi^{(Sl)}_{+}(E,\varphi, \delta E) = i\psi^{(Sl)}_{-}(E,-\varphi,-\delta E), \qquad (l=1,2); \\
  &\mathcal{P}\psi^{(Sl)}_{+}(E) = \psi^{(Sl)}_{-}(-E), \qquad (l=1,2).
\end{align}

Together with the boundary condition
\begin{align}\label{eq:bc}
  &(a_+\psi^{(Ne)}_{+} + a_-\psi^{(Ne)}_{-} +
    b_+\psi^{(Nh)}_{+} + b_-\psi^{(Nh)}_{-})|_{x=0_-} \nonumber\\
  =
  &(c_{1+}\psi^{(S1)}_{+} +
    c_{2+}\psi^{(S2)}_{+} +
    c_{1-}\psi^{(S1)}_{-} +
    c_{2-}\psi^{(S2)}_{-})|_{x=0_+},
\end{align}
where $a_{\pm}$, $b_{\pm}$, $c_{1\pm}$ and $c_{2\pm}$ are constants, one finds the spectrum of the bound states at the interface is given by
\begin{align}\label{eq:ceq}
  [e^{i(\alpha-\beta-2\gamma_-)}-e^{i\varphi}][e^{i(\beta-\alpha-2\gamma_+)}-e^{i\varphi}]=0\,.
\end{align}
This equation can be decoupled to two equations:
\begin{align}
  &F_-=e^{i(\alpha-\beta-2\gamma_-)}-e^{i\varphi}=0, \label{eq:ceq1}\\
  &F_+=e^{i(\beta-\alpha-2\gamma_+)}-e^{i\varphi}=0. \label{eq:ceq2}
\end{align}
They are related by the PHS: $E\rightarrow-E$ ($\alpha\leftrightarrow\beta$, $\gamma_-\leftrightarrow\gamma_+$); \textbf{or}, by the TRS: $\varphi\rightarrow-\varphi$, $\delta E\rightarrow-\delta E$ ($\gamma_-\rightarrow\pi-\gamma_+$). When $\delta E=0$, the two symmetries together imply $\varphi\rightarrow-\varphi$, $E\rightarrow-E$ for each single equation.

Particularly, for $E=0$ solutions (where $\alpha=\beta$), both Eq.~\eqref{eq:ceq1} and Eq.~\eqref{eq:ceq2} reduce to
\begin{align}\label{eq:sp0S}
  e^{i(\varphi+2\gamma)} = 1,
\end{align}
where $\gamma = \arccos(\delta E/\Delta) \in (0,\pi)$. Expanding Eq.~\eqref{eq:ceq1} and \eqref{eq:ceq2} around the $E=0$ solutions, we find the dependence of the energy splitting on the phase difference $\varphi\rightarrow \pi$:
\begin{align}
E_{\pm}(\varphi)\approx \pm   \left(\frac{\sqrt{m^2-\mu^2}\Delta}{\sqrt{m^2-\mu^2}+\Delta}\right)\left(\frac{\varphi - \pi}{2} - \frac{\delta E}{\Delta}\right),
\end{align}
where we assumed that $|m|>|\mu|$ and $\delta E \ll \Delta$. The numerical solution for the energy spectrum is shown in Fig. [3] in the main text. Using the formula for Josephson current $I(\varphi)=-\frac{2e}{\hbar}\sum_{E_n>0} \frac{\partial E_n(\varphi)}{\partial \varphi},$ one can estimate Josephson current carried by the Majorana modes close to $\varphi=\pi$ to be
\begin{align}
  I_{M}(\varphi\rightarrow \pi)\approx -\frac{e}{\hbar}\left(\frac{\sqrt{m^2-\mu^2}\Delta}{\sqrt{m^2-\mu^2}+\Delta}\right),
\end{align}
which dependents on the QPC confinement.



\subsection{Majorana bound states}
To find the wavefunctions of the bound states, we write Eq.~\eqref{eq:bc} explicitly as
\begin{align}
  \Psi(x=0_-) =
  \begin{pmatrix}
	A_+ e^{-i\alpha/2}\\
	A_- e^{i\alpha/2}\\
	A_- e^{-i\alpha/2}\\
	A_+ e^{i\alpha/2}\\
	B_+ e^{-i\beta/2}\\
	B_- e^{i\beta/2}\\
	B_- e^{-i\beta/2}\\
	B_+ e^{i\beta/2}
  \end{pmatrix} =
  \begin{pmatrix}
	c_{1+} e^{i(\varphi+\gamma_+)/2}\\
	c_{1-} e^{i(\varphi+\gamma_-)/2}\\
	c_{2+} e^{-i\gamma_-/2}\\
	c_{2-} e^{-i\gamma_+/2}\\
	c_{1+} e^{-i(\varphi+\gamma_+)/2}\\
	-c_{1-} e^{-i(\varphi+\gamma_-)/2}\\
	-c_{2+} e^{i\gamma_-/2}\\
	c_{2-} e^{i\gamma_+/2}
  \end{pmatrix} =
  \Psi(x=0_+),
\end{align}
where $A_+ = a_++a_-$, $A_- = a_+-a_-$, $B_+ = b_++b_-$ and $B_- = b_+-b_-$. This equation can be broken down into two copies related by PHS:
\begin{align}
  &e^{i(\beta-\alpha-2\gamma_+)}-e^{i\varphi}=0,\quad
  \begin{pmatrix}
	A_+\\
	B_+\\
	c_{1+}\\
	c_{2-}
  \end{pmatrix} =
  \begin{pmatrix}
	e^{i(\varphi+\gamma_+)/2}\\
	e^{i(\beta-\alpha-\varphi-\gamma_+)/2}\\
	e^{-i\alpha/2}\\
	e^{i(\alpha+\varphi+2\gamma_+)/2}
  \end{pmatrix}; \\
  &e^{i(\alpha-\beta-2\gamma_-)}-e^{i\varphi} = 0,\quad
  \begin{pmatrix}
	A_-\\
	B_-\\
	c_{1-}\\
	c_{2+}
  \end{pmatrix} =
  \begin{pmatrix}
	e^{i(\varphi+\gamma_-)/2}\\
	-e^{i(\alpha-\beta-\varphi-\gamma_-)/2}\\
	e^{i\alpha/2}\\
	e^{i(\varphi+2\gamma_- -\alpha)/2}
  \end{pmatrix}.
\end{align}
In particular, for $E=0$ ($\alpha=\beta$, $\gamma_+=\gamma_-=\gamma$), the two degenerate solutions are
\begin{align}
  &e^{i(\varphi+2\gamma)}=1,\quad
  \begin{pmatrix}
	A_+\\
	B_+\\
	c_{1+}\\
	c_{2-}
  \end{pmatrix} =
  \begin{pmatrix}
	e^{-i\gamma/2}\\
	e^{i\gamma/2}\\
	s e^{-i\alpha/2}\\
	e^{i\alpha/2}
  \end{pmatrix},\quad
  \begin{pmatrix}
	A_-\\
	B_-\\
	c_{1-}\\
	c_{2+}
  \end{pmatrix} =
  \begin{pmatrix}
	e^{-i\gamma/2}\\
	-e^{i\gamma/2}\\
	s e^{i\alpha/2}\\
	e^{-i\alpha/2}
  \end{pmatrix},
\end{align}
where $s = e^{i(\varphi+2\gamma)/2} = \pm 1$ owing to Eq.~\eqref{eq:sp0S}. The corresponding wavefunctions are
\begin{align}
  &{\Psi_1}{(x\le 0)} = D e^{(m \sin\alpha)x}
  \begin{pmatrix}
	e^{-i\frac{\gamma}{2}}\cos fx \\
	-i e^{-i\frac{\gamma}{2}}\sin fx \\
	-i e^{-i\frac{\gamma}{2}}\sin fx \\
	e^{-i\frac{\gamma}{2}}\cos fx \\
	e^{i\frac{\gamma}{2}}\cos fx \\
	-i e^{i\frac{\gamma}{2}}\sin fx \\
	-i e^{i\frac{\gamma}{2}}\sin fx \\
	e^{i\frac{\gamma}{2}}\cos fx
  \end{pmatrix}, \;
  {\Psi_1}{(x\ge 0)} = D e^{-(\Delta \sin\gamma)x}
  \begin{pmatrix}
	e^{-i\frac{\gamma}{2}}e^{i(\mu-\delta E)x} \\
	0 \\
	0 \\
	e^{-i\frac{\gamma}{2}}e^{-i(\mu-\delta E)x} \\
	e^{i\frac{\gamma}{2}}e^{i(\mu+\delta E)x} \\
	0 \\
	0 \\
	e^{i\frac{\gamma}{2}}e^{-i(\mu+\delta E)x}
  \end{pmatrix}; \\
  &\Psi_2(x\le 0) = D e^{(m \sin\alpha)x}
  \begin{pmatrix}
	-i e^{-i\frac{\gamma}{2}}\sin fx \\
	e^{-i\frac{\gamma}{2}}\cos fx \\
	e^{-i\frac{\gamma}{2}}\cos fx \\
	-i e^{-i\frac{\gamma}{2}}\sin fx \\
	i e^{i\frac{\gamma}{2}}\sin fx \\
	- e^{i\frac{\gamma}{2}}\cos fx \\
	- e^{i\frac{\gamma}{2}}\cos fx \\
	i e^{i\frac{\gamma}{2}}\sin fx
  \end{pmatrix}, \;
  \Psi_2(x\ge 0) = D e^{-(\Delta \sin\gamma)x}
  \begin{pmatrix}
	0 \\
	e^{-i\frac{\gamma}{2}}e^{-i(\mu+\delta E)x} \\
	e^{-i\frac{\gamma}{2}}e^{i(\mu+\delta E)x} \\
	0 \\
	0 \\
	- e^{i\frac{\gamma}{2}}e^{-i(\mu-\delta E)x} \\
	- e^{i\frac{\gamma}{2}}e^{i(\mu-\delta E)x} \\
	0
  \end{pmatrix}.
\end{align}
Here $D = \text{Diag}(e^{-i\frac{\alpha}{2}},e^{i\frac{\alpha}{2}},e^{-i\frac{\alpha}{2}},e^{i\frac{\alpha}{2}},e^{-i\frac{\alpha}{2}},e^{i\frac{\alpha}{2}},e^{-i\frac{\alpha}{2}},e^{i\frac{\alpha}{2}})$ contains $x$-independent phases that can be absorbed into the definition of the basis.
One verifies that $\mathcal{P}\Psi_1 = \Psi_2$, $\mathcal{T}\Psi_1(\delta E) = -i\Psi_2(-\delta E)$ and $\mathcal{T}\Psi_2(\delta E) = i\Psi_1(-\delta E)$. Therefore we can define two Majorana solutions
\begin{align}
  \chi_1 = e^{i\pi/4}\Psi_1 + e^{-i\pi/4}\Psi_2, \; \chi_2 = i(e^{i\pi/4}\Psi_1 - e^{-i\pi/4}\Psi_2),
\end{align}
which clearly satisfy
\begin{align}
  &\mathcal{P}\chi_1=\chi_1, \; \mathcal{P}\chi_2=\chi_2; \\
  &\mathcal{T}\chi_1(\delta E)=\chi_2(-\delta E),
  \;\mathcal{T}\chi_2(\delta E)=-\chi_1(-\delta E).
\end{align}
Note that $\delta E$ physically represents electro-magnetic vector potential, and hence is reversed under $\mathcal{T}$. When $\delta E \rightarrow 0$, the Majorana solutions are given explicitly by (up to a normalization constant)
\begin{align}
  &{\chi_1}{(x\le 0)} = D e^{\sqrt{m^2-\mu^2}x}
  \begin{pmatrix}
	\cos fx - \sin fx \\
	-i (\cos fx + \sin fx) \\
	-i (\cos fx + \sin fx) \\
	\cos fx - \sin fx \\
	i (\cos fx + \sin fx) \\
	-(\cos fx - \sin fx) \\
	-(\cos fx - \sin fx) \\
	i (\cos fx + \sin fx)
  \end{pmatrix}, \;
  {\chi_1}{(x\ge 0)} = D e^{-\Delta x}
  \begin{pmatrix}
	e^{i\mu x} \\
	-i e^{-i\mu x} \\
	-i e^{i\mu x} \\
	e^{-i\mu x} \\
	i e^{i\mu x} \\
	-e^{-i\mu x} \\
	-e^{i\mu x} \\
	i e^{-i\mu x}
  \end{pmatrix}; \\
  &{\chi_2}{(x\le 0)} = D e^{\sqrt{m^2-\mu^2}x}
  \begin{pmatrix}
	i (\cos fx + \sin fx) \\
	- (\cos fx - \sin fx) \\
	- (\cos fx - \sin fx) \\
	i (\cos fx + \sin fx) \\
	- (\cos fx - \sin fx) \\
	i (\cos fx + \sin fx) \\
	i (\cos fx + \sin fx) \\
	- (\cos fx - \sin fx)
  \end{pmatrix}, \;
  {\chi_2}{(x\ge 0)} = D e^{-\Delta x}
  \begin{pmatrix}
	i e^{i\mu x} \\
	-e^{-i\mu x} \\
	-e^{i\mu x} \\
	i e^{-i\mu x} \\
	-e^{i\mu x} \\
	ie^{-i\mu x} \\
	ie^{i\mu x} \\
	-e^{-i\mu x}
  \end{pmatrix}.
\end{align}

\section{Details of transport calculations}

\subsection{Scattering matrix}
We first derive symmetry properties of the scattering matrix starting from its definition
\begin{align}\label{eq:smat0}
  S_{n'\nu',n\nu}(E) = v_F \int dt\, e^{iEt/\hbar} \theta(t) \bigl\langle\{ \psi_{Ln'\nu'}(x'_0,t), \psi_{Rn\nu}^{\dagger}(x_0,0) \} \bigr\rangle
\end{align}
where $n,n'=1,2$ stand for the lateral edges, $\nu,\nu'=e,h$ stand for electron or hole channels, and by definition $\psi_{e} = \psi$ and $\psi_{h} = \psi^\dagger$ in terms of the original field operators. Both $x'_0$ and $x_0$ must be far away from the scattering region (the QPC in our case). As long as this condition is met, we can conveniently choose $x'_0=x_0$ and drop $x_0$ in our following derivation. Eq.~\eqref{eq:smat0} hence becomes
\begin{align}
  S_{n'\nu',n\nu}(E) = v_F \int dt\, e^{iEt/\hbar} \theta(t) \bigl\langle\{ e^{iHt} \psi_{Ln'\nu'} e^{-iHt}, \psi_{Rn\nu}^{\dagger} \} \bigr\rangle.
\end{align}

By PHS, we have
\begin{align}
  P\psi_{R/Ln\nu}P^{-1} = \psi_{R/Ln\nu}^\dagger = \psi_{R/Ln\bar{\nu}},\;
  P\psi_{R/Ln\nu}^\dagger P^{-1} = \psi_{R/Ln\nu} = \psi_{R/Ln\bar{\nu}}^\dagger\;,
\end{align}
where we have denoted $\bar{e}=h$ and $\bar{h}=e$. Therefore
\begin{align}
  \bigl\langle e^{iHt} \psi_{Ln'\nu'} e^{-iHt} \psi_{Rn\nu}^{\dagger} \bigr\rangle &= \bigl\langle e^{iHt} P \psi_{Ln'\nu'}^\dagger P^{-1} e^{-iHt} P \psi_{Rn\nu} P^{-1} \bigr\rangle \\
  &= \bigl\langle e^{iHt} U_P (\psi_{Ln'\nu'}^\dagger e^{-iHt} \psi_{Rn\nu})^* U_P^\dagger \bigr\rangle \\
  &= \bigl\langle U_P (\psi_{Rn\nu}^\dagger e^{iHt} \psi_{Ln'\nu'})^* U_P^\dagger e^{-iHt} \bigr\rangle^* \\
  &= \bigl\langle P \psi_{Rn\nu}^\dagger e^{iHt} \psi_{Ln'\nu'} P^{-1} e^{-iHt} \bigr\rangle^* \\
  &= \bigl\langle \psi_{Rn\bar{\nu}}^\dagger e^{iHt} \psi_{Ln'\bar{\nu'}} e^{-iHt} \bigr\rangle^*,
\end{align}
where we have used the explicit form $P=U_P K$ with $U_P$ unitary and $K$ the complex conjugate operator, as well as the fact that $P H P^{-1} = -H$. It follows immediately that
\begin{align}
  \bigl\langle\{ e^{iHt} \psi_{Ln'\nu'} e^{-iHt}, \psi_{Rn\nu}^{\dagger} \} \bigr\rangle = \bigl\langle\{ e^{iHt} \psi_{Ln'\bar{\nu'}} e^{-iHt}, \psi_{Rn\bar{\nu}}^{\dagger} \} \bigr\rangle^*,
\end{align}
thus
\begin{align}\label{eq:Sphs}
  S_{n'\nu',n\nu}(E) = S_{n'\bar{\nu'},n\bar{\nu}}(-E)^*, \qquad (\text{PHS}).
\end{align}

By TRS, we have
\begin{align}
  T\psi_{Rn\nu}T^{-1} = s_n\psi_{Ln\nu},\; T\psi_{Ln\nu}T^{-1} = -s_n \psi_{Rn\nu}\;,
\end{align}
where $s_{1}$=$1$ and $s_{2}$=$-1$. Therefore
\begin{align}
  \bigl\langle e^{iHt} \psi_{Ln'\nu'} e^{-iHt} \psi_{Rn\nu}^{\dagger} \bigr\rangle &= -s_ns_{n'} \bigl\langle e^{iHt} T \psi_{Rn'\nu'} T^{-1} e^{-iHt} T \psi_{Ln\nu}^{\dagger} T^{-1} \bigr\rangle \\
  &= -s_ns_{n'} \bigl\langle e^{iHt} U_T (\psi_{Rn'\nu'} e^{i\bar{H}t} \psi_{Ln\nu}^{\dagger})^* U_T^\dagger \bigr\rangle \\
  &= -s_ns_{n'} \bigl\langle U_T (\psi_{Ln\nu} e^{-i\bar{H}t} \psi_{Rn'\nu'}^{\dagger})^* U_T^\dagger e^{-iHt} \bigr\rangle^* \\
  &= -s_ns_{n'} \frac{\text{Tr}[U_T^\dagger e^{-iHt} e^{-\beta H} U_T (\psi_{Ln\nu} e^{-i\bar{H}t} \psi_{Rn'\nu'}^{\dagger})^*]^*}{\text{Tr}  (e^{-\beta H})^*} \\
  &= -s_ns_{n'} \frac{\text{Tr}[T^{-1} e^{-\beta H} e^{-iHt} T \psi_{Ln\nu} e^{-i\bar{H}t} \psi_{Rn'\nu'}^{\dagger}]}{\text{Tr}  (Te^{-\beta H}T^{-1})} \\
  &= -s_ns_{n'} \frac{\text{Tr}[e^{-\beta \bar{H}} e^{i\bar{H}t} \psi_{Ln\nu} e^{-i\bar{H}t} \psi_{Rn'\nu'}^{\dagger}]}{\text{Tr}  e^{-\beta \bar{H}}} \\
  &= -s_ns_{n'} \bigl\langle e^{i\bar{H}t} \psi_{Ln\nu} e^{-i\bar{H}t} \psi_{Rn'\nu'}^{\dagger} \bigr\rangle_{H\rightarrow\bar{H}}\;,
\end{align}
where we have used the explicit form $T=U_T K$ with $U_T$ unitary and $K$ the complex conjugate operator, and have denoted $\bar{H} = T H T^{-1}$. By the same token, we have
\begin{align}
  \bigl\langle \psi_{Rn\nu}^{\dagger} e^{iHt} \psi_{Ln'\nu'} e^{-iHt} \bigr\rangle = -s_ns_{n'} \bigl\langle \psi_{Rn'\nu'}^{\dagger} e^{i\bar{H}t} \psi_{Ln\nu} e^{-i\bar{H}t} \bigr\rangle_{H\rightarrow\bar{H}}\;.
\end{align}
It follows that
\begin{align}\label{eq:Strs}
  S_{n'\nu',n\nu}(E,H) =  -s_ns_{n'} S_{n\nu,n'\nu'}(E,\bar{H}), \qquad (\text{TRS}).
\end{align}
In our case $\bar{H}(\varphi,\delta E) = H(-\varphi,-\delta E)$ where $H$ is the second quantized Hamiltonian, therefore $\bar{H}\ne H$ unless $\delta E=0$ and $\varphi=n\pi$ with $n$ an integer. If we assume $\delta E\approx 0$ and $\varphi\approx n\pi$, then $\bar{H} \approx H$. If we further focus on the regime $E \approx 0$, then  Eqs.~\eqref{eq:Sphs} and \eqref{eq:Strs} together imply
\begin{align}\label{eq:Smat0}
  S =
  \begin{pmatrix}
    0 & b & -ia_1 & c \\
    b & 0 & c^* & -ia_2 \\
    ia_1 & c^* & 0 & b^* \\
    c & ia_2 & b^* & 0
  \end{pmatrix},\quad
  (a_1,a_2\in\mathds{R},\;b,c\in\mathds{C}),
\end{align}
where both incoming and outgoing bases are ordered as (1e, 2e, 1h, 2h). The physical meanings of the scattering amplitudes are clear: $b$ stands for normal back-scattering from one edge to the other; $a_1$ and $a_2$ stand for local Andreev reflections involving only one single edge; $c$ stands for crossed Andreev reflection involving both edges. By further using the unitarity of $S$, we obtain the following equations
\begin{align}
  &(a_1+a_2)b=0,\\
  &(a_1-a_2)c=0,\\
  &bc=0,\\
  &a_1^2+|b|^2+|c|^2=a_2^2+|b|^2+|c|^2=1.
\end{align}
The solutions are limited to the following cases:
\begin{description}
\item[Case 0] $b=c=0$, $a_1,a_2=\pm 1$. This is the case of a completely open QPC;
\item[Case 1] $b \ne 0$, $c = 0$, $a_1=-a_2$. This is the case when the phase difference between two SCs is approximately $2l\pi$. In this case, only local Anreev reflection and normal backscattering (to the other edge) will occur;
\item[Case 2] $c \ne 0$, $b = 0$, $a_1=a_2$. This is the case when the phase difference between two SCs is $(2l+1)\pi$. In this case, only Anreev reflections (local and crossed) will occur; normal backscattering is not possible between two edges.
\end{description}

To see why this is so, it is helpful to start from the limit where the QPC is completely open, namely, $m=f=0$ in the original Hamiltonian. In this limit, the Hamiltonian \eqref{eq:haml0} on the normal side is diagonal: the two edges are completely decoupled, therefore $b=c=0$; we can work on each pairing channel separately. Using the two components $\psi_{R1}$ and $\psi_{L1}^\dag$ as an example, the general solution on the normal side is trivial and the general solution on the SC side is given by Eq.~\eqref{eq:S1sol}. Thus the boundary condition at $x=0$ is given by (assuming $\delta E=0$ and $E=0$)
\begin{align}
  \begin{pmatrix}
	a_R \\
	a_L
  \end{pmatrix} =
  \begin{pmatrix}
    e^{i(\varphi+\pi/2)/2} \\
    e^{-i(\varphi+\pi/2)/2}
  \end{pmatrix}.
\end{align}
The corresponding scattering matrix element is $ia_1 = a_L/a_R = -i e^{-i\varphi}$, that is, $a_1 = -e^{-i\varphi}$. Similarly, working on the two components $\psi_{R2}$ and $\psi_{L2}^\dag$, we find $a_2 = 1$. Clearly, $a_1/a_2=-1$ if $\varphi=2l\pi$; $a_1/a_2=1$ if $\varphi=(2l+1)\pi$. Note that the minus sign of $a_1/a_2$ when $\varphi=2l\pi$ comes from the different ways we impose TRS on the two edges (cf. how $s_n$ is defined); what is more important here is the sign change of $a_1/a_2$ when $\varphi$ changes from $2l\pi$ to $(2l+1)\pi$. Next if we gradually narrow the QPC constriction, while keeping constant $\varphi = n\pi$, we know from the possible cases listed above that $a_1/a_2$ must equal either $+1$ or $-1$ and this sign cannot change abruptly (unless $a_1 = a_2 = 0$, which generically does not happen). It follows that Case 1 must correspond to $\varphi=2l\pi$, whereas Case 2 must correspond to $\varphi=(2l+1)\pi$.

\subsection{Current correlations}
We now focus on the zero-frequency current correlation function defined by
\begin{align}
  P_{12} = \int_{-\infty}^{\infty}dt\, \frac{1}{2} \langle\{ \delta\hat{I}_1(t),\delta\hat{I}_2(0)\}\rangle,
\end{align}
where $\delta\hat{I}_{n = 1,2}(t) = \hat{I}_n(t) - \langle \hat{I}_{n} \rangle$ with $\langle \hat{I}_{n} \rangle$ being the average current. Using the scattering theory for phase coherent transport, $P_{12}$ can be expressed in terms of the scattering matrix and Fermi distribution functions \cite{anantram_current_1996, li_scattering_2012}. In the zero-temperature limit, one finds
\begin{align}
  &P_{12} = -\frac{e^2}{h}\int_{E\geq 0} dE \sum\limits_{n,n'=1,2}\mathrm{Tr}[F_n S_{1n}^{\dagger}\sigma_z S_{1n'} F_{n'} S_{2n'}^{\dagger}\sigma_z S_{2n}]\,, \label{eq:P}\\
  &S_{nn'} =
  \begin{pmatrix}
    S_{ne,n'e} & S_{ne,n'h} \\
    S_{nh,n'e} & S_{nh,n'h}
  \end{pmatrix},\;
  F_n =
  \begin{pmatrix}
    \theta(eV_n-E) & 0\\
    0 & \theta(-eV_n-E)
  \end{pmatrix}.\label{eq:SF}
\end{align}
Here, $n$ and $n'$ are indices for the contacts. This is solely shot noise with two types of contributions: the $n=n'$ terms (in the summation) are partition noise, which will be indicated by a superscript $^{(par.)}$; the $n\ne n'$ terms are exchange noise, which will be indicated by a superscript $^{(ex.)}$. Note that, in principle, Eq.~\eqref{eq:P} must also include contact 3 into the summation for the setup we are considering. But because contact 3 is always grounded ($eV_3=0$) in our discussion, $F_3$ is always a null matrix, and we immediately see that contact 3 is dropped out from Eq.~\eqref{eq:P}.

We are particularly interested in two cases: the equal bias case ($eV_1 = eV_2 = eV>0$; the corresponding correlator is denoted by $P_{12,+}$), and the opposite bias case ($eV_1 = -eV_2 = eV>0$; the corresponding correlator is denoted by $P_{12,-}$). Here, we assume in either case the bias voltage is small enough such that in the relevant range of energy $(0\le E \le eV)$, the scattering matrix can be taken to be energy independent and the constraint Eq.~\eqref{eq:Sphs} from PHS applies:
\begin{align}
  S_{nn'} =
  \begin{pmatrix}
    b_{nn'} & a_{nn}^* \\
    a_{nn} & b_{nn'}^*
  \end{pmatrix}.
\end{align}
For later convenience we denote [cf. Eq.~\eqref{eq:SF}]
\begin{align}
  F_e =
  \begin{pmatrix}
    1 & 0\\
    0 & 0
  \end{pmatrix},\quad
  F_h =
  \begin{pmatrix}
    0 & 0\\
    0 & 1
  \end{pmatrix}.
\end{align}
\begin{description}
\item[Equal bias case:] The partition noise
\begin{align}
  P_{12,+}^{(par.)} &= -\frac{e^2}{h} \left(\int_{0}^{eV} dE\right) \sum\limits_{n=1,2}\mathrm{Tr}[F_e S_{1n}^{\dagger}\sigma_z S_{1n} F_e S_{2n}^{\dagger}\sigma_z S_{2n}] \\
  &= -\frac{e^3V}{h} \sum\limits_{n=1,2}\mathrm{Tr}[F_e S_{1n}^{\dagger}\sigma_z S_{1n} F_e S_{2n}^{\dagger}\sigma_z S_{2n}] \\
  &= -\frac{e^3V}{h} \left[(B_{11}-A_{11})(B_{21}-A_{21})+(B_{12}-A_{12})(B_{22}-A_{22})\right], \label{eq:Peqb_par}
\end{align}
where $B_{nn'} = |b_{nn'}|^2$ and $A_{nn'} = |a_{nn'}|^2$.\\
The exchange noise
\begin{align}
  P_{12,+}^{(ex.)} &= -\frac{e^3V}{h} \mathrm{Tr}[F_e S_{11}^{\dagger}\sigma_z S_{12} F_e S_{22}^{\dagger}\sigma_z S_{21}] + c.c. \\
  &= -\frac{e^3V}{h} (b_{11}^*b_{12}-a_{11}^*a_{12})(b_{22}^*b_{21}-a_{22}^*a_{21})+c.c.\;.
\end{align}
If in addition we are limited to the time-reversal invariant points, where the scattering matrix takes the form of Eq.~\eqref{eq:Smat0}, we obtain
\begin{align}
  &P_{12,+}^{(par.)}
  = \frac{e^3V}{h} (a_1^2+a_2^2)(|b|^2-|c|^2) = 2\frac{e^3V}{h} a_1^2(|b|^2-|c|^2), \\
  &P_{12,+}^{(ex.)}
  = 2\frac{e^3V}{h} a_{1}a_{2}|c|^2 = 2\frac{e^3V}{h} a_{1}^2|c|^2,\\
  &P_{12,+}=P_{12,+}^{(par.)}+P_{12,+}^{(ex.)}
  = 2\frac{e^3V}{h} a_{1}^2|b|^2\,.
\end{align}
\item[Opposite bias case:] The partition noise
\begin{align}
  P_{12,-}^{(par.)} &= -\frac{e^3V}{h} \mathrm{Tr}[F_e S_{11}^{\dagger}\sigma_z S_{11} F_e S_{21}^{\dagger}\sigma_z S_{21} + F_h S_{12}^{\dagger}\sigma_z S_{12} F_h S_{22}^{\dagger}\sigma_z S_{22}] \\
  &= -\frac{e^3V}{h} \left[(B_{11}-A_{11})(B_{21}-A_{21})+(B_{12}-A_{12})(B_{22}-A_{22})\right] \\
  &= P_{12,+}^{(par.)} = P_{12}^{(par.)}\,.
\end{align}
Here, $P_{12,-}^{(par.)} = P_{12,+}^{(par.)}$ is a result of PHS.\\
The exchange noise is
\begin{align}
  P_{12,-}^{(ex.)} &= -\frac{e^3V}{h} \mathrm{Tr}[F_e S_{11}^{\dagger}\sigma_z S_{12} F_h S_{22}^{\dagger}\sigma_z S_{21}] + c.c. \\
  &= -\frac{e^3V}{h} (b_{11}^*a_{12}^*-a_{11}^*b_{12}^*)(a_{22}b_{21}-b_{22}a_{21})+c.c.\;.
\end{align}
If in addition the scattering matrix takes the form of Eq.~\eqref{eq:Smat0} with time-reversal invariance, we obtain
\begin{align}
  &P_{12,-}^{(par.)}
  = \frac{e^3V}{h} (a_1^2+a_2^2)(|b|^2-|c|^2) = 2\frac{e^3V}{h} a_1^2(|b|^2-|c|^2), \\
  &P_{12,-}^{(ex.)}
  = 2\frac{e^3V}{h} a_{1}a_{2}|b|^2 = -2\frac{e^3V}{h} a_{1}^2|b|^2,\\
  &P_{12,-}=P_{12,-}^{(par.)}+P_{12,-}^{(ex.)}
  = -2\frac{e^3V}{h} a_{1}^2|c|^2\,.
\end{align}
\end{description}
It is interesting to notice that, indeed,
\begin{align}
  P_{12}^{(par.)}+P_{12,+}^{(ex.)}+P_{12,-}^{(ex.)} = P_{12,+}+P_{12,-}-P_{12}^{(par.)} = 0.
\end{align}
This result is not limited to the time-reversal invariant points (although it becomes more transparent at these points), but can be proved for all $\varphi$ by using the unitarity of the scattering matrix $S$.

\subsection{Numerical simulations}
\begin{figure}
  \centering
  \includegraphics[width=0.85\textwidth]{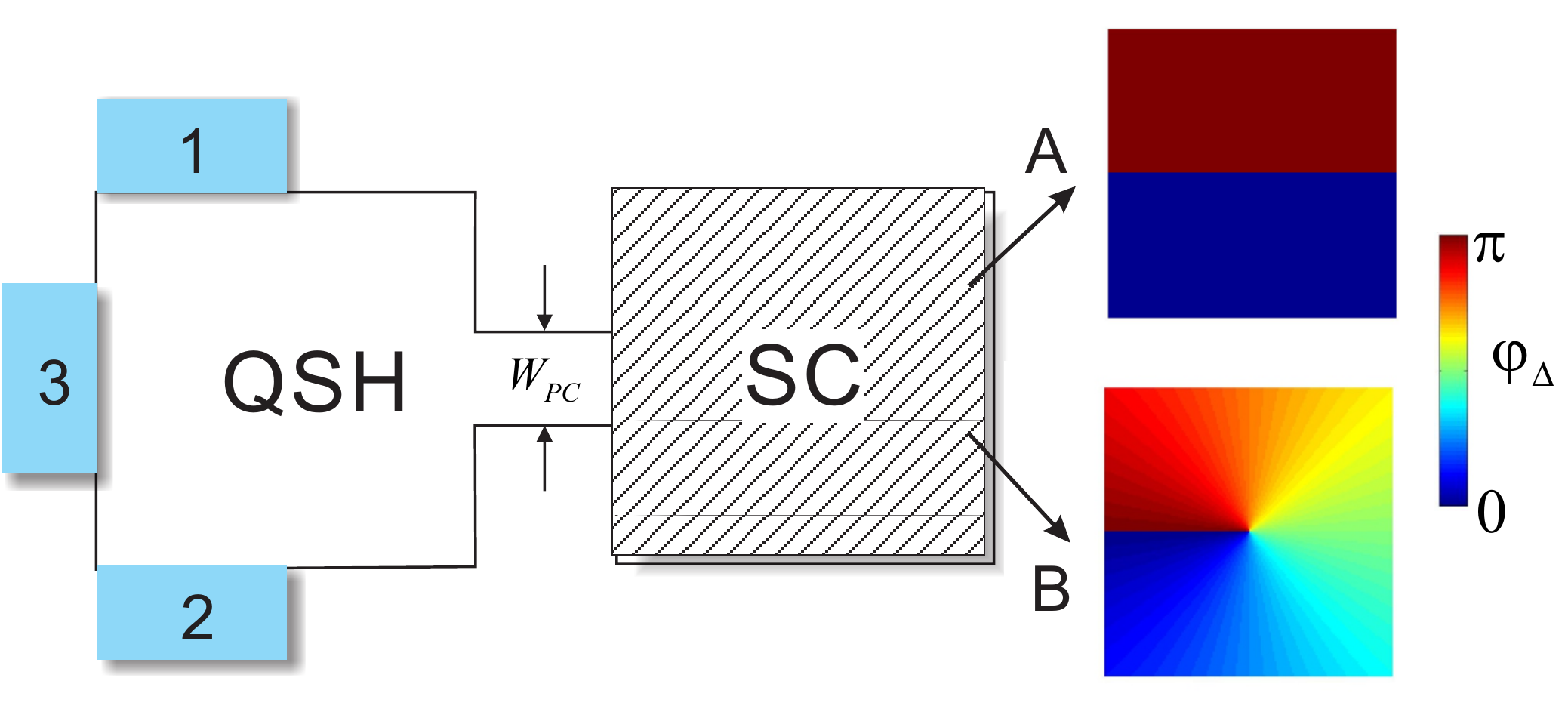}
  \caption{A sketch of the simulation setup. Depending on the size of the covering superconductor, there are two scenarios for the Josephson junction: in scenario A, we assume the pairing phases to be constant on either side of the junction; in scenario B, we consider a slow spatial variation of the pairing phases. The color-coded profiles for the pairing phases in both scenarios are shown here. The lattice size ($N_x\times N_y$) used in our simulations is: 160$\times$120 for the left (uncovered) block; 20$\times W_{PC}$ for the constriction with $W_{PC}$ kept as a variable; 100$\times$120 for the right (SC-covered) block unless otherwise specified.}
  \label{fig:setup_sim}
\end{figure}
A sketch of the simulation setup is shown in Fig.~\ref{fig:setup_sim}. It is the same as that in Fig. 1 of the main text except for being regularized with right-angle corners. We assume the Bernevig-Hughes-Zhang Hamiltonian \cite{bernevig06} for the underlying QSH sample. On a square lattice, this Hamiltonian reads
\begin{align}
  H_{BHZ} &= \sum\limits_{i_x,i_y} \left[c_{i_x,i_y}^\dagger T_0 c_{i_x,i_y} - (c_{i_x+1,i_y}^\dagger T_x c_{i_x,i_y} + c_{i_x,i_y+1}^\dagger T_y c_{i_x,i_y} + h.c.)\right], \\
  T_0 &= (C+4D)\sigma_0\otimes\tau_0 + (M+4B)\sigma_0\otimes\tau_3 + \Delta_{BIA} \sigma_2\otimes\tau_2\,, \\
  T_x &= D\sigma_0\otimes\tau_0 + B\sigma_0\otimes\tau_3 + (A/2i)\sigma_3\otimes\tau_1\,, \\
  T_y &= D\sigma_0\otimes\tau_0 + B\sigma_0\otimes\tau_3 + (A/2i)\sigma_0\otimes\tau_2\,,
\end{align}
where $c_{i_x,i_y}$ is a (column) four-vector, $\sigma$'s and $\tau$'s are Pauli matrices (including identity) for spin and orbital degrees of freedom, respectively. Note that we have included an additional spin-orbit coupling term, proportional to $\Delta_{BIA}$, which can be associated with bulk inversion asymmetry \cite{konig08}. The purpose of including this term is to simulate generic situations where spin is not a good quantum number. Also, since our simulations concern only generic (qualitative rather than quantitative) features related to our proposed setup, we will use simplified (and hypothetical) parameters instead of the parameters extracted experimentally for specific materials/devices. The parameters used are: $C=D=0$, $A=B=1$, $M=-0.2$ and $\Delta_{BIA}=0.05$. In addition, the magnitude of the superconducting proximity gap is chosen to be $\Delta = 0.05$. Although the values of the parameters are not realistic, we have checked that varying these parameters does not change the qualitative features predicted in this paper, and hence we expect our predictions to hold in real experiments.

\subsubsection{Effects of varying QPC width and SC width}
\begin{figure}
\begin{center}
\begin{minipage}{\textwidth}
  \centering
  \includegraphics[width=0.65\linewidth]{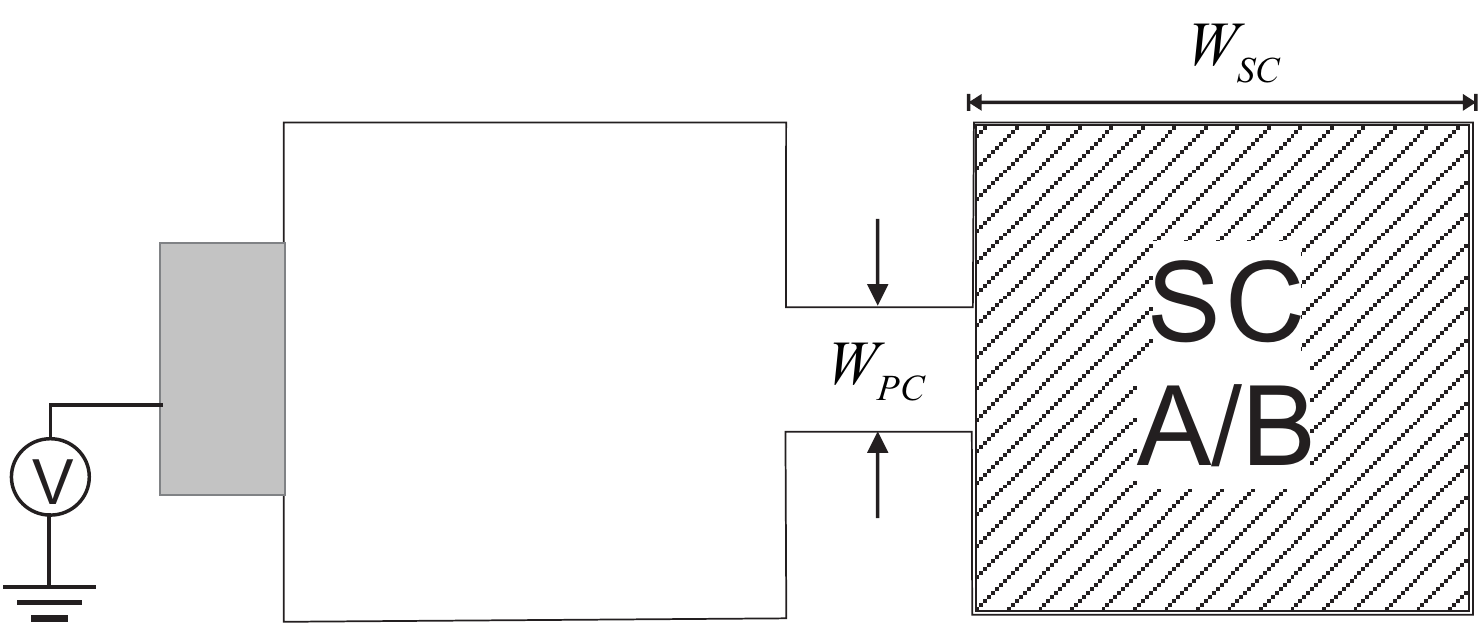}
\end{minipage}
\end{center}
\begin{center}
\begin{minipage}{\textwidth}
  \centering
  \includegraphics[width=\linewidth]{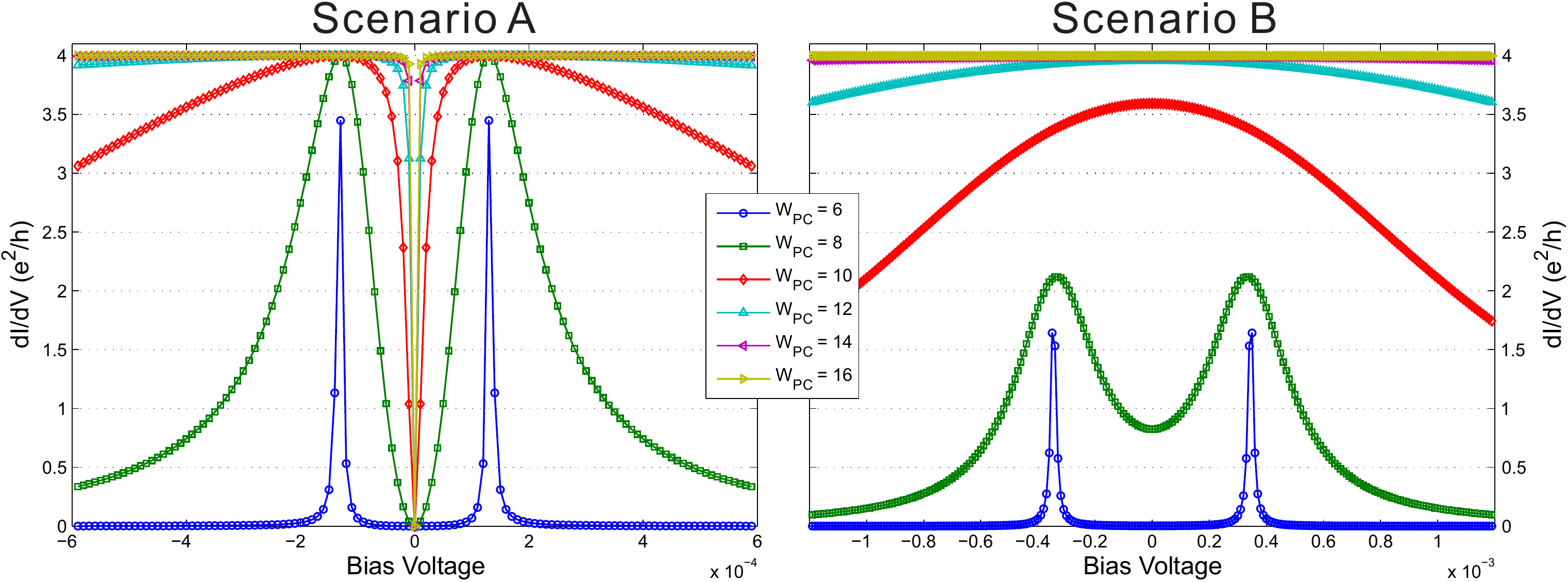}
\end{minipage}
\end{center}
  \caption{Differential conductances measured around zero bias voltage in a single-metallic-contact setup with a fixed SC width $W_{SC}=80$. The two scenarios are compared with various QPC width. In both scenarios we set $\varphi = \pi$; in scenario B we set additionally $\epsilon_0=0.016/rad$, where $\epsilon_0$ is the slope of the superconductivity order parameter phase with respect to the polar angle near the superconductor junction (cf. Fig.~\ref{fig:setup_sim}).}
  \label{fig:c1}
\end{figure}

\begin{figure}
\begin{center}
  \centering
  \includegraphics[width=\linewidth]{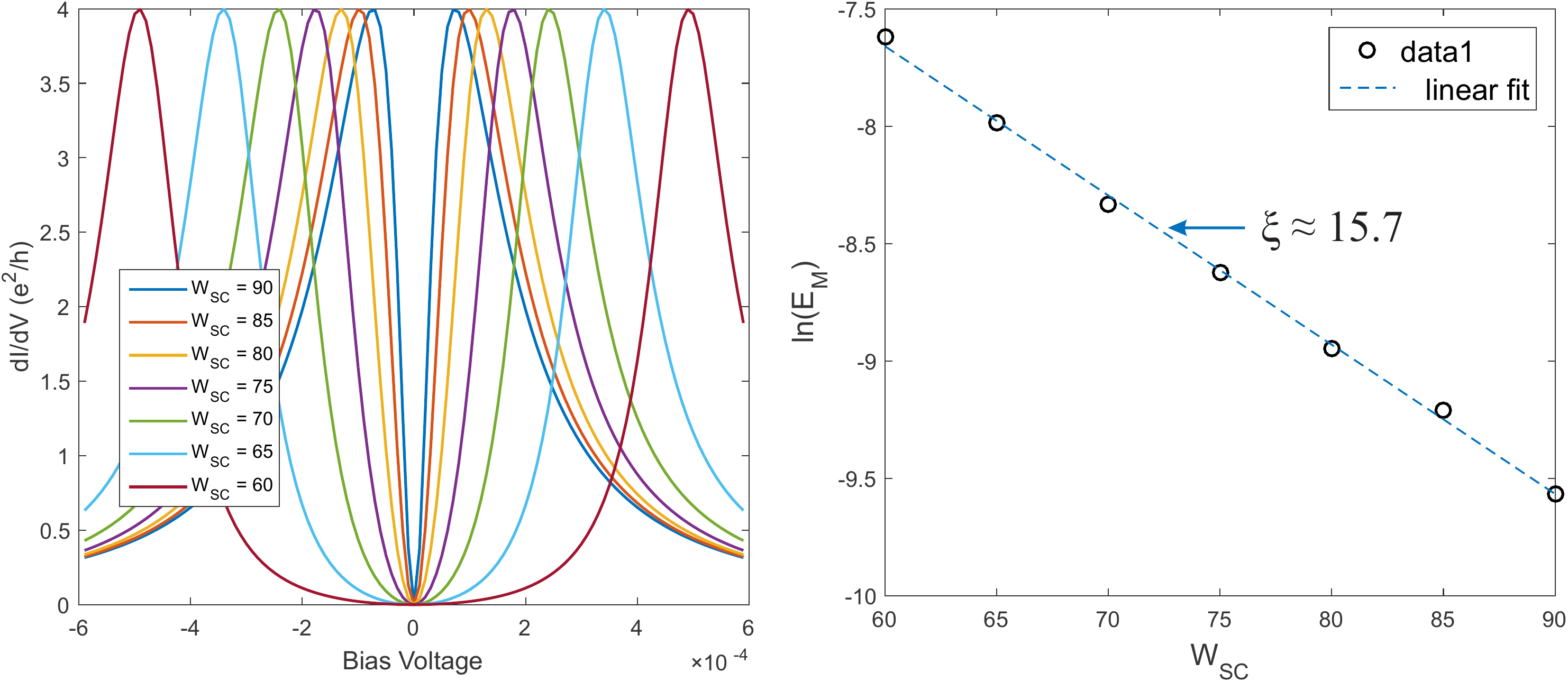}
\end{center}
  \caption{Differential conductances (left panel) measured at nearly-zero energy in a single-metallic-contact setup with various width $W_{SC}$ for the superconductor-covered region in Scenario A ($W_{PC} = 8$; see also Fig.~\ref{fig:c1}). The double peaks in the differential conductances correspond to the splitting energy $E_M$ of the Majorana bound states due to the finite size effects. Indeed, there is another Majorana Kramers pair at the opposite ends of a superconductor which has a finite overlap with Majorana modes localized at QPC. It is expected that the splitting energy should decay exponentially with $W_{SC}$. By fitting the splitting energy with respect to $W_{SC}$ (right panel), we extract the effective superconducting coherence length to be about $15.7$ (in units of lattice constant) in our simulations.}
  \label{fig:Wsc}
\end{figure}

In Fig.~\ref{fig:c1} we compare the differential conductances measured in a single-metallic-contact setup (which is equivalent to $2G_1$, as in Eq.~(21) of the main text, in our originally proposed multi-terminal setup) with various QPC width in two scenarios (see Fig.~\ref{fig:c1} caption). We notice that the differential conductances in general exhibit a double-peak structure with respect to the bias voltage because of a finite splitting energy of coupled MBSs. The peaks are broadened when the QPC is gradually softened. In scenario A, as a consequence of strict TRS, the occurrence of a Majorana Kramers pair at QPC-Josephson junction interface is always accompanied by the occurrence of another pair at the far end of the Josephson junction. The coupling of these MBSs across the SC-covered sample leads to an exponentially small splitting energy. The precise dependence of this splitting energy on the sample size has been investigated in Fig.~\ref{fig:Wsc}, where the differential conductances with various width of SC-covered region has been compared. This allows us to extract the superconducting coherence length in our simulations to be $\xi\simeq15.7$ in units of lattice constant. Despite the splitting, the peak value of difference conductance in scenario A is always $4e^2/h$, corresponding to perfect Andreev reflections induced by the Majorana pair. In scenario B, where TRS applies only approximately, the peak value can be lower than $4e^2/h$, but approaching this value with sufficiently wide QPCs. Note that the quantized peak value at $4e^2/h$ here implies that $G_{21}$ in our originally proposed multi-terminal setup is exclusively given by crossed Andreev reflections [see analysis below Eq.~(21) in the main text].

\subsubsection{Comparison of different cases for the SC-covered region}
\begin{figure}
\begin{center}
\begin{minipage}{\textwidth}
  \centering
  \includegraphics[width=\linewidth]{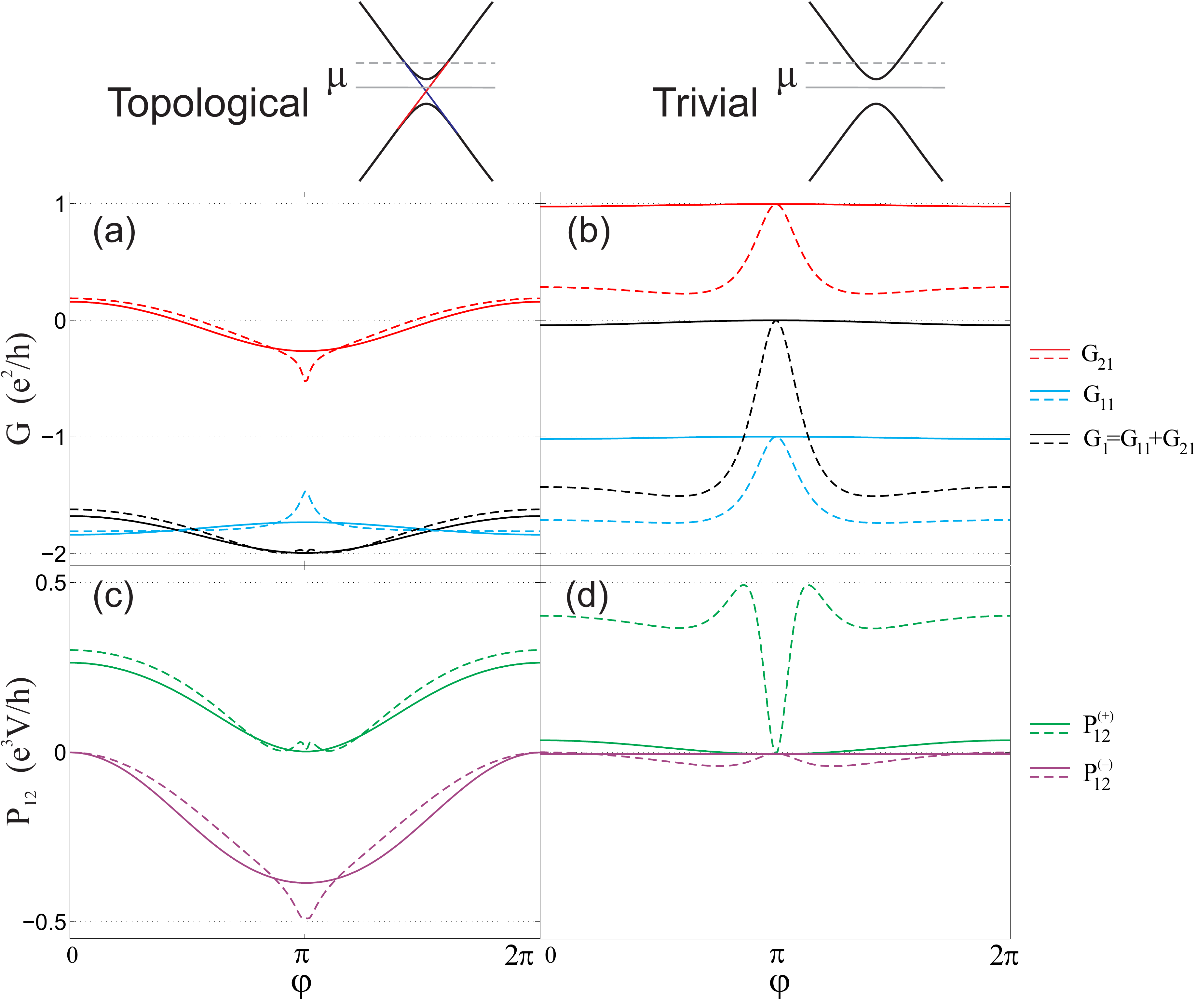}
\end{minipage}
\end{center}
  \caption{Typical conductance $G$ and current cross-correlation $P_{12}$ when the SC-covered region is topological [panels (a) and (c); $M=-0.2$] or trivial [panels (b) and (d); $M=0.2$], and when the chemical potential at the SC-covered region lies in the bulk gap (solid lines; $\mu=0$) or traverses bulk bands (dashed lines; $\mu=0.4$). The width of the QPC $W_{PC} = 14$ in all cases. Note that in Fig.~4 of the main text, $W_{PC} = 10$, which is different from here, but the main features are the same.}
  \label{fig:c2}
\end{figure}
To complement the results shown in Fig.~4 of the main text, here in Fig.~\ref{fig:c2} we compare cases where the SC-covered region is topological (left panels) or trivial (right panels), as well as cases where the chemical potential at the SC-covered region lies in the bulk gap (solid lines) or traverses bulk bands (dashed lines). The conclusions from this comparison is the following: in topologically nontrivial (QSH) samples, the occurrence of Majorana bound states at $\varphi=(2n+1)\pi$ always leads to an enhanced crossed Andreev reflection (see the red curves in Fig.~\ref{fig:c2} upper-left panel), whereas in trivial samples at the same $\varphi$ the crossed Andreev reflection is always suppressed (see the red curves in Fig.~\ref{fig:c2} upper-right panel); in terms of the total one-terminal conductance $G_1$ at $\varphi=(2n+1)\pi$, the former implies a strong total (Andreev) conductance ($G_1\simeq -2e^2/h$; see the black curves in Fig.~\ref{fig:c2} upper-left panel) and the latter implies a weak total conductance ($G_1\simeq 0$ because of the cancellation between the Andreev and normal reflections; see the black curves in Fig.~\ref{fig:c2} upper-right panel); in terms of the current cross-correlations, the former implies a strong negative correlation (because of the almost-exclusive Andreev processes; see the purple curves in Fig.~\ref{fig:c2} lower-left panel) when the two contacts are oppositely biased, whereas the latter implies a suppressed such correlation (see the purple curves in Fig.~\ref{fig:c2} lower-right panel).

\end{widetext}


\end{document}